\documentclass[a4paper,11pt,leqno]{article}
\usepackage{amsfonts}     
\usepackage{amsmath}
\usepackage{amstext}
\usepackage{amsthm}             
\usepackage{xspace}
\usepackage[dvips]{graphicx}

\newcommand\ac{\`a\xspace}

\newcommand{\N}{\mathbb N}

\newtheorem{theorem*}{Theorem}
\newtheorem{prop*} {Proposition} 
\newtheorem{lemma*}{Lemma}
\newtheorem{lemma}{Lemma}

\theoremstyle{definition}
\newtheorem{definition}{Definition}
\newtheorem{definition*}{Definition}
\newtheorem{cor*}{Corollary}
\newtheorem{rem*}{Remark}
\theoremstyle{remark}

\newtheorem{dim*}{\bf Proof}
   
\newtheorem{guess*}{\bf Osservazione}
  
\input psfig.sty
\begin{document}
\title{Sublinear growth of Information in DNA sequences}
\author{Giulia Menconi\\ \small {Dipartimento di Matematica
Applicata}\\ \small{and}\\ \small {C.I.S.S.C. Centro
Interdisciplinare} \\ \small {per lo Studio dei Sistemi Complessi}\\
\small {Universit\ac di Pisa}\\ \small {Via Bonanno Pisano 25/b 56126
PISA - Italy}\\ \small{menconi@mail.dm.unipi.it}\\October 23, 2003} \date{} \maketitle
\vskip 11truecm
\centerline{Running title: Sublinear Information in DNA}
\newpage
\begin{abstract}
We introduce a novel method to analyse complete genomes and recognise
some distinctive features by means of an adaptive compression
algorithm, which is not DNA-oriented. We study the Information Content
as a function of the number of symbols encoded by the
algorithm. Preliminar results are shown concerning regions having a
sublinear type of information growth, which is strictly connected to
the presence of highly repetitive subregions that might be supposed to
have a regulatory function within the genome.
\end{abstract}
\section{Introduction}
We shall analyse the genome sequences from the point of view of data
compression in order to exploit a linguistic analysis. As the context
suggests, the genomes are interpreted as symbol sequences of finite
length, drawn by an Information Source (the Nature) that remains
mainly unknown and emits symbols taken from the alphabet of the four
nucleotides $\{A,\ C,\ G,\ T\}$. Each genome identifies a living
organism and we assume that it may be considered as the unique
realisation produced by the Source relative to that organism.

We shall not give here a formal definition of Information
Source. Intuitively, it is a device emitting a sequence of symbols
$\dots x_1x_2x_3\dots$ where each $x_i$ is an element of a finite
alphabet $\mathcal A$. The rigorous definition \cite{billingsley} lies
on the notion of sequence space $\Omega_\mathcal A$, that is the space
of one-sided infinite sequences (also called strings)
$\omega=(\omega_0,\omega_1,\dots)$ whose symbols are drawn from the
alphabet.
Even if an Information Source is rigorously defined as a stochastic
process $\mathbb X=(\mathbb X_n)_{n\in\N}$ acting on a sequence space,
we may consider the symbolic source $\Omega _{\mathcal A}$ as the
subset of the sequence space containing all the realizations of the
process $\mathbb X$. This shall motivate the use of the term
Information Source when referring to a sequence space. We shall denote
by $\mathcal{A}^*$ the set of finite symbolic sequences on the
alphabet $\mathcal{A}$. If $s\in\mathcal{A}^*$ its length will be
denoted by $|s|$.

DNA sequences are special quaternary symbol sequences. As only a small
fraction of DNA nucleotides results in a viable organism, the
sequences belonging to a living organism are expected to be nonrandom
and have some constraints. Therefore, DNA sequences should be
compressible, at least locally.

In our approach to symbol sequences, the crucial notion is the
\textit{Information Content}. Given a finite string $s$ in $\mathcal{A}^*$,
the meaning of \textit{ quantity of information} $I(s)$ contained in
$s$ has the following natural connotation:

\begin{center}
$I(s)$ \textit{is the length of the smallest binary message from which you
can reconstruct} $s$.
\end{center}

In his pioneering work, Shannon defined the quantity of information as
a statistical notion using the tools of probability theory
(\cite{kin}). Thus in Shannon framework, the quantity of information
which is contained in a string depends on its context. For example the
string $^{\prime }pane^{\prime }$ contains a certain information when
it is considered as a string coming from the English language. The
same string $^{\prime }pane^{\prime }$ contains much less Shannon
information when it is considered as a string coming from the Italian
language because it is more frequent in the Italian language (in
Italian it means ''bread'' and, of course, it is very
frequent). Roughly speaking, the Shannon information of a string is
the absolute value of the logarithm of the probability of its
occurrence.

However, there are measures of information which depend intrinsically
on the string and not on its probability within a given context. We
will adopt this point of view. An example of these measures of
information is the Algorithmic Information Content ($AIC$). We will
not formally define it (see \cite{kin} and \cite{Ch} for rigorous
definitions and properties). We limit ourselves to give an intuitive
idea which is very close to the formal definition. We can consider a
partial recursive function as a computer $C$ which takes a program $p$
(namely a binary string) as an input, performs some computations and
gives a string $s=C(p)$, written in the given alphabet, as an output.
The $AIC$ of a string $s$ is defined as the length of the shortest
binary program $p$ which gives $s$ as its output, namely
$$I_{AIC}(s,C)=\min \{|p|:C(p)=s\}, $$ where $|p|$ means the length in
bit of the string which the program $p$ consists of. A theorem due to
A. N. Kolmogorov (\cite{kolmogorov}) implies that the information
content ${AIC}$ of $s$ with respect to $C$ depends only on $s$ up to a
fixed constant, therefore its asymptotic behaviour does not depend on
the choice of $C$. The shortest program $p$ which outputs the string
$s$ is a sort of optimal encoding of $s$. The information that is
necessary to reconstruct the string is contained in the
program. Unfortunately, this coding procedure cannot be performed by
any algorithm. This is a very deep statement and, in some sense, it is
equivalent to the Turing halting problem or to the G\"{o}del
incompleteness theorem. Then the Algorithmic Information Content is
not computable by any algorithm.

Our method is focused on another measure: the information content of a
finite string can also be defined by a lossless data compression
algorithm $Z$ (\cite{Ch}, \cite{cleary}). This turns out to be a
Computable Information Content (CIC). In reference \cite{licatone}
quantitative relations among Shannon entropy of the source, the AIC
and the CIC of sequences are provided.

The ``classical'' studies in compression algorithms answer the
question about the com\-pres\-si\-bi\-li\-ty of DNA with the
additional advantage of using compression techniques to capture the
properties of DNA. It is known that DNA sequences have two linguistic
characteristic structures: {\it reverse complements} and {\it
approximate repeats}. The reverse complement $\sigma ^c$ of a sequence
$\sigma$ is a sequence such that each symbol of $\sigma$ is replaced
in $\sigma^c$ by its complement one. That is, reading the reverse
complement of a subsequence from a single strand of DNA is the same as
reading the corresponding complementary subsequence in the other
strand. The approximate repeats are repeats that contain
errors. Approximate repeats are due to the local variability that is a
common feature within genomes.

There have been developed several special-purpose compression
algorithms for DNA sequences (for instance, see  \cite{cleary}, \cite{jiang},
\cite{chen}, \cite{tahi}). These
algorithms are called DNA-oriented because they use the
aforementioned charateristic structures of ge\-no\-mes together with a
sort of statistical compression to achieve a compression ratio lower
than two bits per symbol. This is a great improvement since the
standard text compression algorithms such as {\it compress} or {\it
gzip} cannot compress DNA sequences but only expand the file with more
than two bits per symbol. The reason for text compression to fail on
DNA sequences is that the regularities in genomes are much more
subtler than in English texts, for which those algorithms have been
designed.

Our analysis makes reference to a different approach. We aim at using
the compression algorithm CASToRe, which has been created without any
biological purpose and {\it a priori} linguistic knowledge, to
understand whether there exist low information regions within a
genome, whether they have a functional type in common, whether they
are extended or have short length and what kind of growth the
information content shows in those regions.
Finally, as the algorithm CASToRe belongs to the class of algorithms
that adaptively create a dictionary relative to a parsing of the input
sequence, we shall study dictionaries after compression, in order to
investigate the relations between patterns and biological functions.
\section{Computable Information Content}
\begin{definition}[Compression Algorithm]
A lossless data compression algorithm is any injective function
$Z:\mathcal{A}^*\rightarrow\{0,1\}^*$.
\end{definition}
Therefore, a compression algorithm is a reversible coding such that
from the original string $s$ may be recovered from the encoded string
$Z(s)$. Since the coded string contains all the information that is
necessary to reconstruct and describe the structural features of the
original string, we can consider the length of the coded string as an
approximate measure of the quantity of information that is contained
in the original string.
\begin{definition}[Computable Information Content]
The information content of a finite string $s\in:\mathcal{A}^*$ with
respect to a compression algorithm $Z$ is defined as
\begin{equation}
CIC_{Z}(s)=|Z(s)|\ .
\end{equation}
The CIC of a string $s$ is the length (in bit units) of the coded
string $Z(s)$.
\end{definition}
The advantage of using a compression algorithm lies in the fact that
the information content $CIC_{Z}\left( s\right) $ is a
computable function over the space of finite strings. For this reason
we named it Computable Information Content.

Moreover, we define another quantity, the complexity of a finite
sequence, providing an estimate for the rate of information content
contained in it.

\begin{definition}[Computable Complexity of a finite string]
The complexity of $s$ with respect to $Z$ is the compression ratio
\begin{equation}
K_{Z}(s)=\frac{I_Z(s)}{|s|}\ .
\end{equation}
\end{definition}

\begin{rem*}
Under suitable optimality assumptions on the compression algorithm
$Z$, we can extend this definition to infinite symbolic sequences
belonging to $\Omega_\mathcal A$ and asympotically obtain the Shannon
entropy of the Information Source from which the sequence has been
drawn (\cite{gal4},\cite{gal3}). The theoretical work
has been extended also to trajectories coming from general dynamical
systems and it is supported by application to several complex systems,
as to turbulent or intermittent regimes (\cite{CSF02}, \cite{giuliauno},
\cite{bonanno}, \cite{cristalli}, \cite{jacopogiulia}) and to weakly
chaotic dynamical systems (\cite{menconi},\cite{licatone}).
\end{rem*}

\section{Dictionaries, words and phrases}
Let us describe the sort of linguistic analysis we shall perform on
 genetic sequences. We shall use the CIC method to extract the
 functional regions whose information content is low and its growth is
 sublinear. We aim at understanding whether those regions show
 peculiar features such as specific highly repeated patterns of
 nucleotides (they are usually called {\it motifs}). Finally, we shall
 scan other genomes, both coming from the same domain of life and from
 different domains, looking for the presence of low information
 regions and comparing the motifs to each other. These regions are
 called {\it atypical}, as surprisingly they are highly compressible
 in comparison with the other regions. The dictionaries of some
 atypical regions will be studied and related to some known biological
 functions (e.g. being a promoter region). Finally, a preliminar result on
 potential application of this method to gene finding will be
 introduced.

\subsection{The algorithm CASToRe}\label{castore}
We have created and implemented a particular compression algorithm we
called CASToRe which is a modification of the Lempel-Ziv compression
schemes $LZ77$ and $LZ78$ (\cite{lz77}, \cite{lz78}) and it has been
introduced and studied in references \cite{CSF02} and
\cite{menconi}. Its theoretical advantages with respect to LZ78 showed
that this algorithm is a sensitive measure of the Information content
of low entropy sequences. This is the reason that motivates the choice
of the acronym \textbf{CASToRe} to name the new algorithm: its meaning
is \textbf{C}ompression \textbf{A}lgorithm, \textbf{ S}ensitive
\textbf{To} \textbf{Re}gularity. As it has been proved in
\cite{menconi}, the Information content $I_Z$ of a constant sequence
$s^n$, originally with length $n$, is $\Psi(n)=4+2\log (n+1)[\log (\log
(n+1))-1]$, if the algorithm $Z$ is CASToRe. The theory predicts that
the best possible information content for a constant sequence of
length $n$ is $AIC(s^n) =\log (n) + $constant. It may be shown that
the algorithm $LZ78$ encodes a constant $n$-digits long sequence to a
string with length about $const\ +\ n^{\frac 1 2}$ bits; so, we cannot
expect that $LZ78$ is able to distinguish a sequence whose information
content grows like $n^{\alpha}$ ($\alpha < \frac 1 2$) from a constant
or periodic string. Furthermore, the running time of CASToRe is also
sensibly shorter than that of $LZ77$ (with infinite window), then any
implementation is more efficient. These are the main reasons that
motivate the choice of using CASToRe also for numerical experiments.

Now we briefly describe the internal running of CASToRe.

As the Ziv-Lempel schemes, the algorithm CASToRe is based on an
adaptive dictionary (\cite{bell}). One of the basic differences in the
coding procedure is that the algorithm $LZ77$ splits the input strings
in overlapping phrases, while the algorithm CASToRe (as well as the
algorithm $LZ78$) parses the input string in non-overlapping
phrases. Moreover, CASToRe differs from $LZ78$ because the new phrase
is a pair of two already parsed phrases, while $LZ78$ couples one
already parsed phrase and one symbol from the alphabet.
 
At the beginning of encoding procedure, the dictionary contains only
the alphabet. In order to explain the main rules of the encoding, let
us consider a step $h$ within the encoding process, when the dictionary
already contains $h$ phrases $\{e_1,\dots,e_h\}$.
 
The new phrase is defined as a pair ({\it prefix pointer},{\it suffix
pointer}). The two pointers are referred to two (not necessarily
different) phrases $\rho_p$ and $\rho_s$ chosen among the ones
contained in the current dictionary as follows. First, the algorithm
reads the input stream starting from the current position of the front
end, looking for the longest phrase $\rho_p$ matching the
stream. Then, the algorithm looks for the longest phrase $\rho_s$ such
that the joint word $\rho_p+ \rho_s$ matches the stream. The new
phrase $e_{h+1}$ that will be added to the dictionary is then
$e_{h+1}=\rho_p+ \rho_s$.
 
The output file contains an ordered sequence of the binary encoding of 
the pairs $(i_p,i_s)$ such that $i_p$ and $i_s$ are the dictionary 
index numbers corresponding to the prefix word $\rho _p$ and to the 
suffix word $\rho_s$, respectively.  The pair $(i_p,i_s)$ is referred 
to the new encoded phrase $e_{h+1}$ and has its own index number 
$i_{h+1}$. 

\subsubsection{Example}
 
The following example shows how the algorithm CASToRe encodes the 
input stream 
\begin{equation*} 
\omega =(abcababccabb\dots) . 
\end{equation*} 
 
Let the source alphabet be $\mathcal{A}=\{a,b,c\}$. 
 
The output file corresponds to the binary encoding of the following
pairs contained in the second column. The first column is the
dictionary index number of the encoded phrase in the dictionary which
is showed in the same line, second column. For an easier reading, we
add a third column which shows each encoded phrase in the original
stream $\omega$, but which is not contained in the output file:
$$ 
\begin{array}{lll} 
&\mbox{First, the alphabet is loaded}&\\
1 & (0,\ ^{\prime}a\ ^{\prime}\ ) & [a] \\  
2 & (0,\ ^{\prime}b\ ^{\prime}\ ) & [b] \\  
3 & (0,\ ^{\prime}c\ ^{\prime}\ ) & [c] \\  
&\mbox{Then, the encoding procedure starts}&\\
4 & (1,2) & [ab] \\  
5 & (3,4) & [cab] \\  
6 & (4,3) & [abc]\\
7 & (5,3) & [cabc]
\end{array} 
$$ 
and so on. 

\subsection{Reading the dictionary}
The dictionary built by the algorithm CASToRe is an ordered collection
of phrases, that is, of pairs of words. Thus, a phrase is
composed by a prefix-word and a suffix-word. By construction, phrases
are different from each other, since the algorithm exploits a parsing
on the input string. Furthermore, each phrase may become a word, if it
appears as prefix or suffix of other phrases in the following
dictionary.

In the following, we shall look at the most frequent words, at the
longest phrases and in some cases we shall compare the results to the
same analysis performed by means of the algorithm $LZ77$ and exploited
in collaboration with a group of physicists from the University of
Rome (see their previous work \cite{loreto} by V. Loreto et al. for
details on the methodology). We shall show that recurrent subsequences
occur especially along the regions with lowest information
content. Notice that we refer to exact repeats.

We shall distinguish among recurrent subsequences either {\bf
motifs} or {\bf patterns}. A {\it motif} is a recurrent word in the
dictionary, whereas a {\it pattern} is a recurrent subsequence that
does not match any word of the dictionary, but is contained in some of
them. If a motif is found, we shall follow its {\it descent}, that is
the set of phrases whose the motif is either a prefix or a suffix or
both. Moreover, we shall search for the motif to be a {\bf sliding
pattern}, in the sense that it is contained in other phrases
without being their prefix nor their suffix. Furthermore, if only a
sliding pattern is to be found, then we shall recover its {\bf root},
that is the longest word of the dictionary matching part of the
pattern.
\section{The Information Content of DNA sequences}
We have analysed the computable complexity of 12 complete
genomes\footnote{The genomes have been downloaded by means of
the GenBank sequence libraries
http://www.ncbi.nlm.nih.gov/Genbank/index.html} of some Archaea,
Bacteria and Eukaryotes, together with chromosomes II and IV of
\textit{ Arabidopsis thaliana}. The complete list is shown on the
following Table \ref{cssh1}.

In order to take into account the biological
functional constraints actually existing among the bases within the
genome and to highlight new features of coding and noncoding regions, we
have exploited a {\it fragment analysis}.
\begin{definition}
We say that any exon, intron or intergenic region is a functional
{\it fragment} of the genome sequence, following the prediction as it
has been identified via biological databases and statistical tools
(\cite{myers}).
\end{definition}
{\bf Notation. }In prokariotic genomes there are two functional types,
therefore we shall denote by $Coding\_\#$ and $Inter\_\#$ the coding
and the noncoding fragments, respectively, where $\#$ is an index to
order fragments. In eukaryotic genomes there are three different types
of regions: we shall denote by $Exon\_\#$ the coding fragments and by
$Intron\_\#$ and $Inter\_\#$ the noncoding intragenic fragments and the
noncoding integenic fragments, respectively.

Thus, we shall consider the Computable Complexity $K(f)$
of each fragment and study the Information Content growth $CIC(f)$ within a
fragment.

First, we have considered how the Information Content varies along
some complete DNA sequences: that is, we have studied the behaviour of
the CIC of a genome as a function of the number of encoded symbols. As
a result, we remark that the function $CIC(\sigma_n)$ grows linearly for all
the complete genomes $\sigma$ we have analysed and the asymptotic
slope is the value of their Computable complexity $K(\sigma)$:
$$CIC(\sigma_n)\ \sim\ K(\sigma)\cdot n \ ,$$ where $\sigma_n$
indicates the first $n$ bases in the complete genome $\sigma$. However, we can
enhance some regions of the genome and we will see that the $CIC$-line
is locally no more straight. This characteristic feature is shared by
all the genomes we have analysed, both Prokaryotes and
Eukaryotes and confirms the intuitive idea that the Information
Content growth should be slower in the parts of the genome where some
regularity prevails.

\begin{figure}[hb]
\begin{tabular}{lr}
\raggedright{(a)\psfig{figure=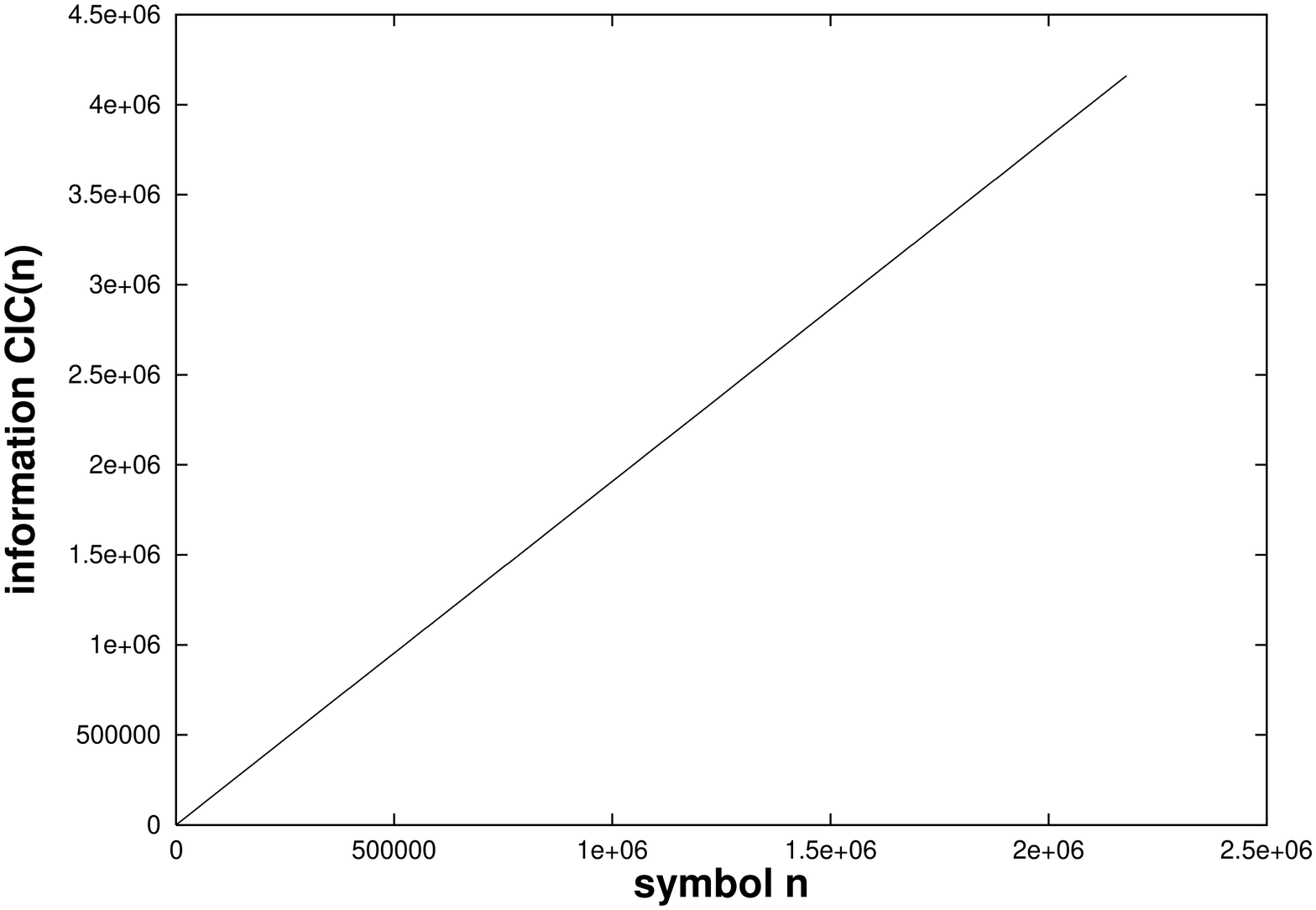,width=6.5cm,angle=270}} &
\raggedleft {(b)\psfig{figure=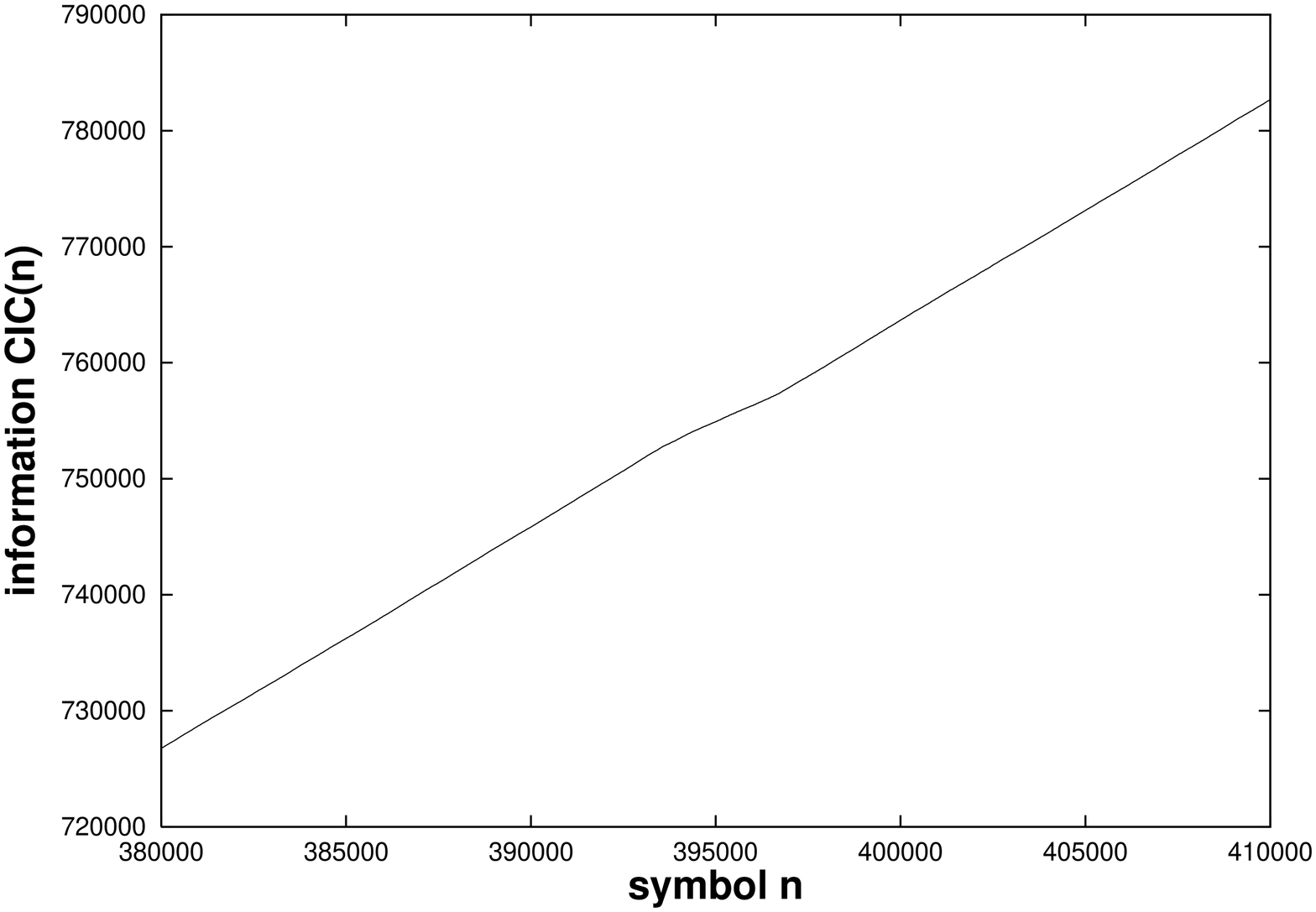,width=6.5cm,angle=270}}
\end{tabular}
\caption{\it (a) complete $CIC(n)$ graph for {\it Archaeoglobus
fulgidus} complete genome; (b) local enhancement of the region from 380000 to
 410000 bp. The behaviour of $CIC(n)$ is no more linear.}\label{cfrTot}
\end{figure}

For instance, see the results about the genome of {\it Archaeoglobus
fulgidus} (Prokaryote) which are pictured on figure \ref{cfrTot}. For
the sake of brevity, we shall not show analogous pictures coming from
other genomes. 

\begin{table}\begin{center}
\begin{tabular}{|c|c|c|}
\hline
\textbf{Genome} & \it{CSS}&{\bf$H_1$} \\ \hline\hline
\textit{Methanococcus jannaschii} & 1.794&1.887 \\ \hline
\textit{Archeoglobus fulgidus} & 1.909&1.987 \\ \hline
\textit{Methanobacterium thermoautrophicum} & 1.907 &1.986\\ \hline
\textit{Pyrococcus abyssi} & 1.901&1.979 \\ \hline\hline
\textit{Aquifex aeolicus} & 1.883 &1.976\\ \hline
\textit{Escherichia coli} & 1.893 &1.987\\ \hline
\textit{Bacillus subtilis} & 1.870 &1.975\\ \hline
\textit{Haemophylus influenzae} & 1.866 &1.947\\ \hline
\textit{Mycoplasma genitalium} & 1.848 &1.959\\ \hline
\textit{Thermotoga maritima} & 1.893 &1.984\\ \hline\hline
\textit{Arabidopsis thaliana} (chr. II and IV) & 1.892&1.938 \\ \hline
\textit{Saccharomyces cerevisiae} & 1.889 &1.949\\ \hline
\textit{Caenorhabditis elegans} & 1.777&1.936 \\ \hline
\end{tabular}
\caption{\it complete genomes. Comparison CSS
vs. $H_1$.}\label{cssh1}\end{center}
\end{table}

For what concerns the values of computable complexity $K$ for the
complete genomes we have analysed, the results are shown on Table
\ref{cssh1} . We have indicated the complexity $K$ as $CSS$, meaning
{\it complexity as a single string}, to distinguish it from the
fragment complexity, which is the value of the computable complexity
of the functional fragments within the complete genome and which will
be denoted by $FC$ in the following. The final column in Table
\ref{cssh1} shows the first order entropy $H_1$ of the sequence. If
$p_A,\ p_C,\ p_G,\ p_T$ are the nucleotide frequencies over a genome
$\sigma$ (the frequency is calculated as the number of occurrences of
a specific nucleotide over the total number of nucleotides), then the
first order entropy is $H_1=\sum_{i=A,C,G,T}p_i\log p_i$. We recall
that, when the symbols are drawn uniformly at random from the source
and all the positions in the sequence are independent from each other,
an optimal coding procedure will devote $\log _{2}(\#\mathcal{A})$
bits per symbol to represent each character (\cite{coverthomas}),
where $\#\mathcal{A}$ is the number of symbols in the alphabet
$\mathcal{A}$. In this case the asymptotically maximal complexity
equals the $H_1$ value for those values of nucleotide frequencies. For
quaternary sequences, like the genomes, this maximal mean first order
entropy is 2 bits per symbol. Since the $H_1$ value represents a
quantity of information of a single string which is dependent on the
probability measure on the space of sequences, at first sight the
genomes cannot be considered randomly distributed (from a statistical
point of view), because for all of them the $H_1$ values are different
from 2 bits per symbol. First, we notice that the values of the
complexity $CSS$ are significantly different from 2 and lower than the
$H_1$ entropy values. Again, this is in complete agreement with the
fact that the randomness of the genomes has strong constraints. It is
also possible to clearly recognise that some genomes have very low
computable complexity (smaller than 1.90 bits per symbol), which means
that their internal structure presents mid-range and long-range
correlations.

The compression of complete genomes does not satisfy the quest for
local structures along a genome. The presence of local nonlinearities
in the Information Content function for complete genomes suggests the
existence of specific functional fragments whose Information Content
function grows sublinearly. We recall that we named those regions
atypical. Consequently, we shall investigate in this direction by
means of the fragment analysis.

\subsection{A sublinearity index}
In order to identify the regions where the growth of the function
$CIC(\sigma_n)$ is sublinear, we define a sublinearity index, that allows us
to determine whether a functional region is atypical. 

In the following, $\sigma$ shall denote any fragment within a
genome. The sublinearity index may be defined by means of any
adaptive compression algorithm $Z$, although the experimental results are
referred to the algorithm CASToRe.

Let $N=|\sigma|$ be the length of the input sequence $\sigma$. Let
$\mathcal{P}(\sigma,Z)$ be the parsing of $\sigma$ with respect to the
algorithm $Z$:
$\mathcal{P}(\sigma,Z)=\{\phi_1,\phi_2,\dots,\phi_t\}$. Therefore, the
input string $\sigma$ is the ordered juxtaposition of phrases $\phi
_j$'s. We use the symbol $n_k$ to indicate the current total number of
encoded symbols up to step $k$ of the encoding procedure:
$n_k=\Sigma_{j=1}^{k}|\phi _j|$. Due to the fact that
$|\phi_k|=n_k-n_{k-1}$, we say that $n_k$ is the parsing index
corresponding to the phrase $\phi _k$. The Information Content after
$k$ steps is then the quantity
$I(n_k)=\Sigma_{j=1}^{k}I(\phi_j)$. Obviously, it holds that
$n_t=\Sigma_{j=1}^{t}|\phi _j|=N$ and
$I(\sigma)=I(N)=\Sigma_{j=1}^{t}I(\phi_j)$. Since the encoding
procedure might be not precise in the early steps as well as in the
final steps, we fix two bounds defining the restriction of the
potential integer value $n_j$. Let $T_{inf}=20\% |\sigma|$ be the
lower bound and $T_{sup}=90\%|\sigma|$ be the upper bound. The choice
of the bounds will be such that there exist two parsing indexes
$n_{inf}$ and $n_{sup}$ such that $T_{inf}\leq n_{inf}<n_{sup}\leq
T_{sup}$. Moreover, since the algorithm $Z$ requires that the input
sequence is sufficiently long to make the compression reliable and
efficient, we shall not analyse sequences whose length $N$ is lower
than $200$ symbols. Thus, for the set $\{n_j\ \|\ j=1,\dots,t\}$
coming from the parsing of $\sigma$ via the algorithm $Z$, we define
the domain $\mathcal{D}=\{n_k\ \|\ n_{inf}\leq n_k\leq n_{sup}\ ,\
n_t\geq 200\}$.

\begin{definition}[Sublinearity index of a finite symbol sequence]\label{sublinind}

$\qquad$\\{\it Let $q_{min}$, $q_{max}$ and $q_Z(\sigma)$ be defined
as follows:
\begin{equation*}
q_{min}=\min\limits_{n_k\in\mathcal{D}}\left\{\frac{I(n_k)}{n_k}\right\}\ ,
\end{equation*}
\begin{equation*}
q_{max}=\max\limits_{n_k\in\mathcal{D}}\left\{\frac{I(n_k)}{n_k}\right\}
\end{equation*}
and
\begin{equation*}
q_{_Z}(\sigma)=\frac{q_{min}}{q_{max}}\ .
\end{equation*}
The sublinearity index $\mathcal{G}_{_Z}(\sigma)$ of the input sequence
$\sigma$ with respect to the parsing defined via the algorithm $Z$ is
the quantity
\begin{equation}\label{gi}\index{$\mathcal{G}_{_Z}$, sublinearity index}
\mathcal{G}_{_Z}(\sigma)=\frac{\log(q_{_Z}(\sigma))}{\log(\frac
{n_{sup}}{n_{inf}})}+1\ .
\end{equation}}
\end{definition}

The definition of this index $\mathcal{G}_{_Z}$ deserves some
comments. Its main characteristic is that it allows a criterion to
identify atypical regions to be established.

First of all, it is known that the behaviour of the Information
Content of a finite sequence $\sigma$ is an increasing function
$I(\sigma^n)$ that grows at most linearly with the number $n$ of
encoded symbols. Therefore, the indexes $q_{min}$ and $q_{max}$ can be
easily calculated by:
$$q_{min}=\frac{I(n_{sup})}{n_{sup}}\ \ \mbox{and}\ \
q_{max}=\frac{I(n_{inf})}{n_{inf}}\ .$$
Hence, it is straightforward that the value of the sublinearity index is 
\begin{equation}\label{utileG}
\mathcal{G}_{_Z}(\sigma)=\frac{\log(I(n_{sup}))-\log(I(n_{inf}))}
{\log(n_{sup})-\log(n_{inf})}\ .
\end{equation}

We notice that the fragment we have analysed are not periodic,
otherwise the phrases found in the parsing by the algorithm CASToRe
would definitely show length doubling, which is absent in the
dictionaries of all the fragments. Furthermore, the Information
Content growth of any functional fragment $\sigma$ can not be a
logarithmic function $\Psi(n)$ (see Section \ref{castore}), but we
might assume that it can be read ($\forall\ 1\leq n\leq |\sigma|$) as
\begin{equation}
CIC(\sigma_n)=\mathcal O(Cn^{\gamma})\ ,\mbox{ with
exponent $0<\gamma\leq 1$ and constant $C>0$}\ .
\label{infoPo}
\end{equation} 

Note that this formula is relative to a finite sequence, therefore the
writing $\mathcal O(Cn^{\gamma})$ is not referring to an asympotic
behaviour (as $n\leq |\sigma|$), but it means that the integer
function $CIC(\sigma_n)$ is fitted by a function whose do\-mi\-nant term is
a power law with exponent smaller than 1. Since we have excluded any
pure periodicity, hypothesis (\ref{infoPo}) is doubtless
plausible. 


The two following main points are definitely true. First, a
sublinear growth of Information Content is an indicator of the
presence of some regularity in the input sequence and this is much
more evident when the index $\mathcal{G}_{_Z}$ is significantly
smaller than 1. Second, small values of the index $\mathcal{G}_{_Z}$
may correspond to different sublinear information growths $-$ also
other than power-law-like $-$ that consequently might be a signal of
different underlying dynamics generating the symbol sequences.

In the following Lemma, the sublinearity index $\mathcal{G}_{_Z}$ in
the case of Information Content growing exactly as a power law is evaluated.
\begin{lemma}
If $CIC(n_k)=C {n_k} ^\gamma$ with $0<\gamma\leq 1$, then
$\mathcal{G}_{_Z}=\gamma$.\end{lemma}
\begin{proof}
 Consider the formula (\ref{utileG}).  In this case, it holds that
$$\mathcal{G}_{_Z}=\frac{\log (C)+\alpha\log(n_{sup})-\log
(C)-\alpha\log(n_{inf})}{\log(n_{sup})-\log (n_{inf})}\ .$$ Therefore,
the conclusion is straightforward.\end{proof}

Thus, according to formula (\ref{infoPo}), the sublinearity index
$\mathcal{G}_{_Z}$ is a reliable quantity that allows the degree of
sublinearity of the information content growth to be estimated. In
order to evaluate the precision of the index $\mathcal{G}_{_Z}$ with
respect to the {\it true} actual exponent $\gamma$, we have compared
the values of $\mathcal{G}_{_Z}$ with the values of $\gamma$ as they
are given by a numerical fit on the integer function $I(n)$. The
results are definitely satisfactory. Some examples are shown on
Table \ref{tabellalfa} and are referred to several fragments from the
genomes of {\it Archaeoglobus fulgidus}, {\it Escherichia coli} and
{\it Arabidopsis thaliana}.

\begin{table}\begin{center}
\begin{tabular}{|c|c|c|c|}
\hline 
\mbox{Genome}&\mbox{Sequence}&\mbox{value of }$\mathcal{G}_{_Z}$
&\mbox{fit-value of }$\gamma$\\ 
\hline\hline
$Archaeoglobus\ fulgidus$&$Coding\_685495$&0.965&1.000\\
\hline
$Archaeoglobus\ fulgidus$&$Inter\_1143603$&0.949&0.949\\
\hline
$Archaeoglobus\ fulgidus$&$Inter\_393196$&0.832&0.831\\
\hline\hline 
$Escherichia\  coli$&$Inter\_2302612$&0.768&0.747\\
\hline 
$Escherichia\  coli$&$Inter\_4293752$&0.728&0.730\\ 
\hline
$Escherichia\  coli$&$Coding\_91419$&0.986&0.986\\
\hline\hline 
$Arabidopsis\ thaliana$&$Exon\_23950656$&0.614&0.585\\
\hline
$Arabidopsis\ thaliana$&$Intron\_5063613$&0.767&0.738\\
\hline
$Arabidopsis\ thaliana$&$Inter\_19660110$&0.887&0.886\\
\hline
\end{tabular}
\caption{\it reliability of the sublinearity index $\mathcal{G}_{_Z}$ in the
case of several functional regions from different genomes.}\label{tabellalfa}
\end{center}
\end{table}

The following definition will be used to extract the atypical
functional regions. The threshold has been fixed according to the
empirical principle that the kind of growth $n^\gamma$ where $\gamma$
lies in $[0.9,1]$ is, on a general basis, equivalent to a linear
growth, due to the finiteness of the sequences under analysis.

\begin{definition}[Atypical region]\label{atypical}
An atypical region within a genome is any functional region whose
sublinearity index $\mathcal{G}_{_Z}$ is smaller than 0.9.
\end{definition}

\begin{figure}
\centerline{\psfig{figure=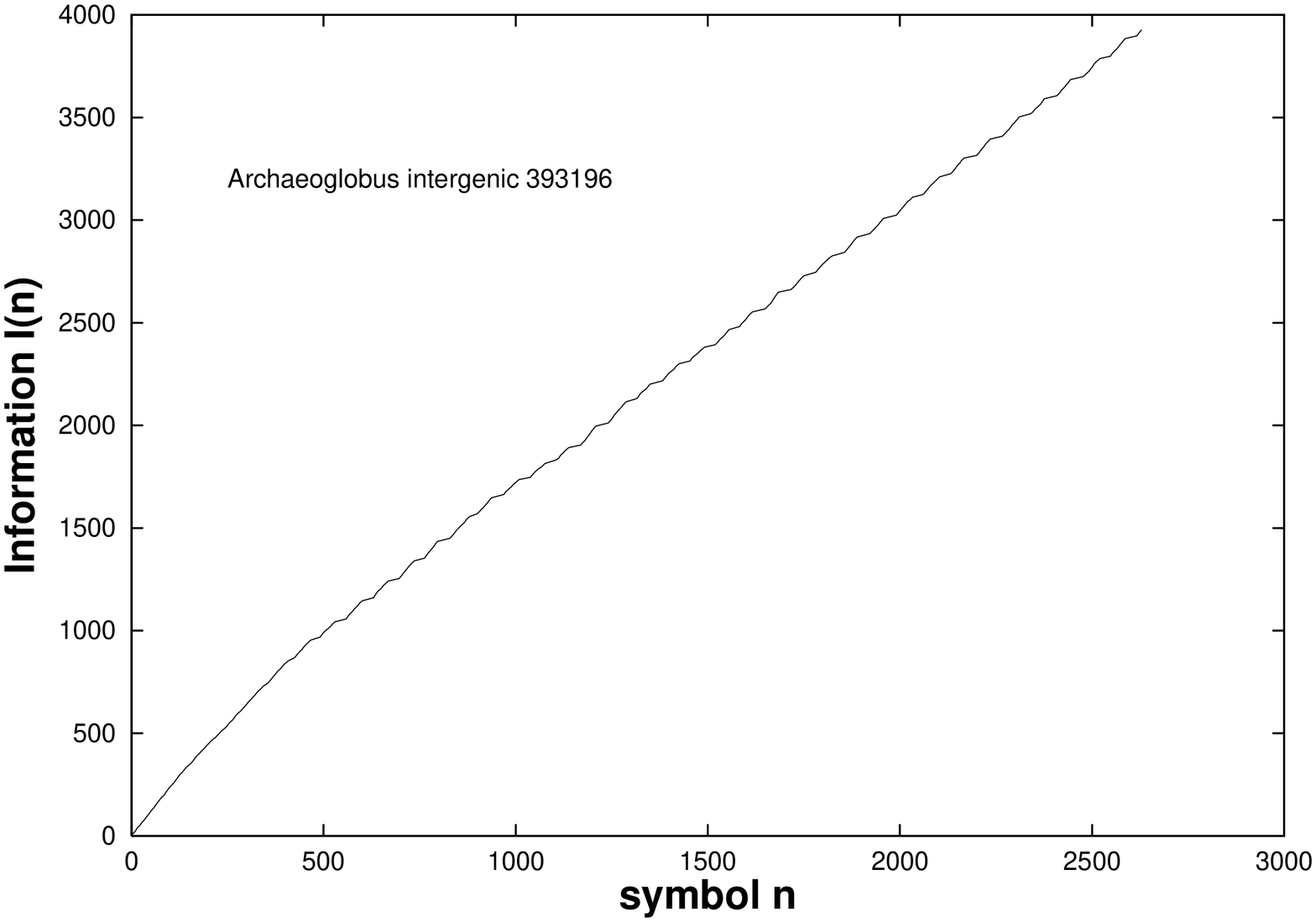,width=8cm,angle=270}}
\caption{\it Archaeoglobus fulgidus genome. The behaviour of the
information content of region $Inter\_393196$ is a power law whose
exponent is 0.832. The picture is in linear scale.}\label{regLow}
\end{figure}
\begin{figure}
\centerline{\psfig{figure=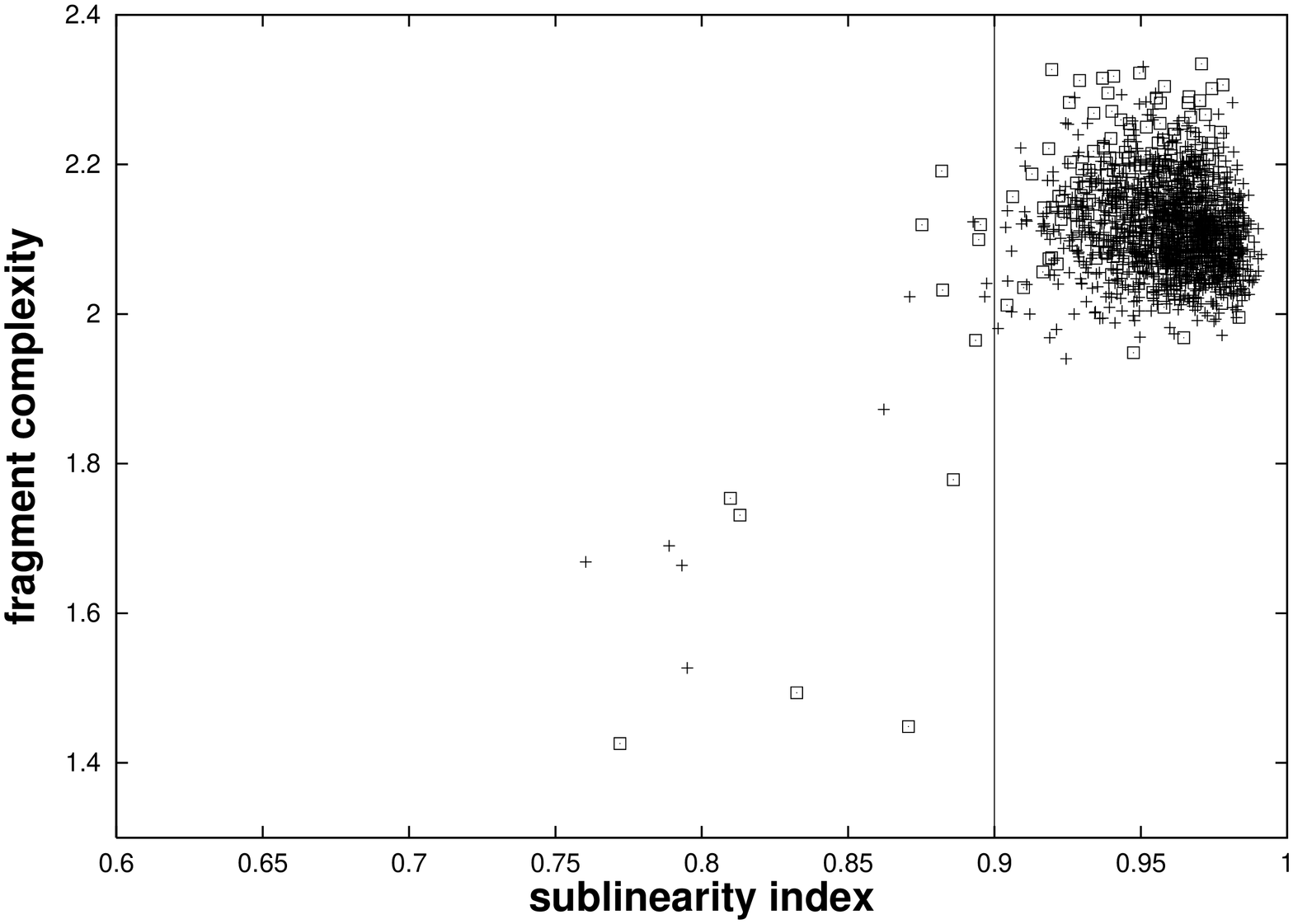,width=8cm,angle=270}}
\caption{\it Archaeoglobus fulgidus genome. Comparison between the
values of sublinearity index and fragment complexity of all functional
regions with length greater than 200 bp. The crosses ($+$) are referred
to coding regions, while the diamonds ($\diamond$) are referred to
intergenic regions. The vertical line is the threshold for the
sublinearity index, under which the region is atypical.}\label{cfrGK}
\end{figure}

The connection between sublinearity index and fragment complexity is
not precise, even if in the extreme cases where both values are either
high or low a sort of clusters are detected. For instance, Figure
\ref{cfrGK} illustrates what the relation is between the sublinearity
index (horizontal axis) and the fragment complexity (vertical axis) in
the case of the genome of {\it Archaeoglobus fulgidus}. Atypical
regions are indicated by means of a vertical line that represents the
threshold for the sublinearity index as introduced in Definition
\ref{atypical}. It is clear that, both in the case of coding regions
(depicted by a cross) and of noncoding regions (depicted by a
diamond), the higher the fragment complexity is, the higher the
sublinearity index is.  Furthermore, the detection of atypical regions
with high fragment complexity suggests that the sublinearity index may
be more meaningful in identifying regularity of sequences than the
fragment complexity.
\section{Experimental results}
In the following, we shall introduce some preliminar examples of
application of the $CIC$ method. Ww shall analyse the dictionary of
some long atypical regions within the genomes of {\it Archaeoglobus
fulgidus}, {\it Methanococcus jannaschii} and {\it Arabidopsis
thaliana}. We shall discover peculiar properties and propose
some biological motivations to those features. This part of the work
has been developed in collaboration to the Animal Biology and Genetics
Department of the University of Florence.

\subsection{Archaeoglobus fulgidus}

{\it Archaeoglobus fulgidus} is a sulphur-metabolizing anaerobic
organism. It belongs to the Archaeoglobales, archaeal sulfate reducers
unrelated to other sulfate reducers. They grow at extremely high
temperatures. Archaeoglobus species causes corrosion of iron and steel
in oil and gas processing systems by the production of iron
sulphide. This organism has one circular chromosome.
\begin{figure}
\centerline{\psfig{figure=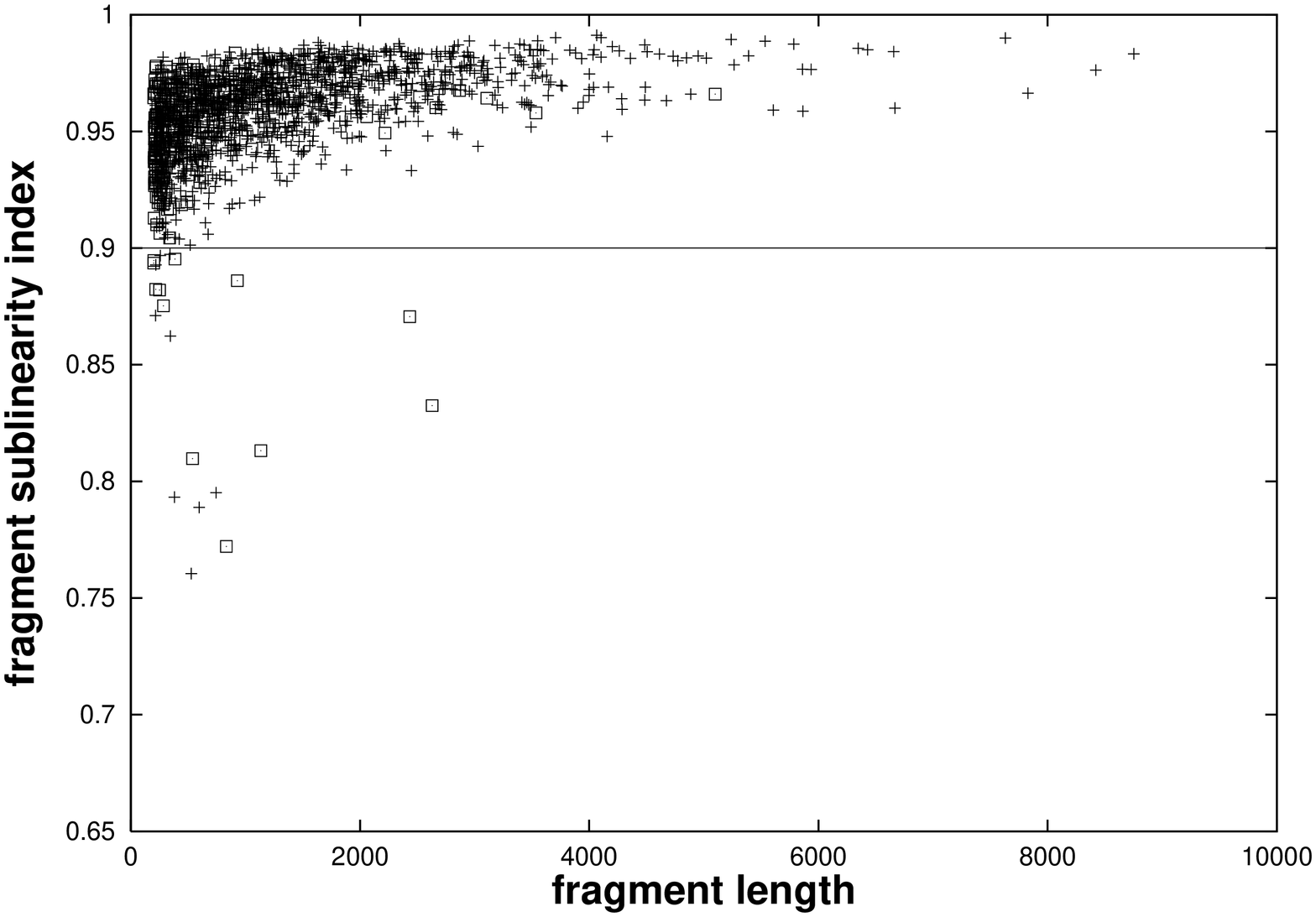,width=8cm,angle=270}}
\caption{\it Archaeoglobus fulgidus genome. Of each functional region,
  its length and the corresponding sublinearity index are
  plotted. The crosses ($+$) are referred to coding regions,
  while the squares ($\square$) are referred to intergenic regions.
  The horizontal line is the threshold for the sublinearity index, under
  which the region is atypical.}\label{cfrLGglobus}
\end{figure}

Looking at Figure \ref{cfrLGglobus}, we have extracted two regions:
one atypical region, which is noncoding, and two non-atypical regions,
one coding and one noncoding. This choice is aimed at comparing the
dictionaries of regions with sublinear grwoth of information to the
dictionaries of regions with li\-near growth of information.

The exemplified regions are
\begin{itemize}
\item $Coding\_685495$: non-atypical region, length $L=2300\ bp$,
  sublinearity index $\mathcal{G}_{_Z}=0.965$, fragment complexity $K=
  2.108$;
\item $Inter\_1143603$: non-atypical region, length $L=2219\ bp$,
 sublinearity index $\mathcal{G}_{_Z}=0.949$, fragment complexity $K= 2.117$;
\item $Inter\_393196$: atypical region, length $L=2629\ bp$,
  sublinearity index \linebreak$\mathcal{G}_{_Z}=0.832$, fragment
  complexity $K= 1.494$.
\end{itemize}
We start analysing the non-typical regions.
\begin{figure}
\begin{tabular}{lr}
\raggedright{(a)\psfig{figure=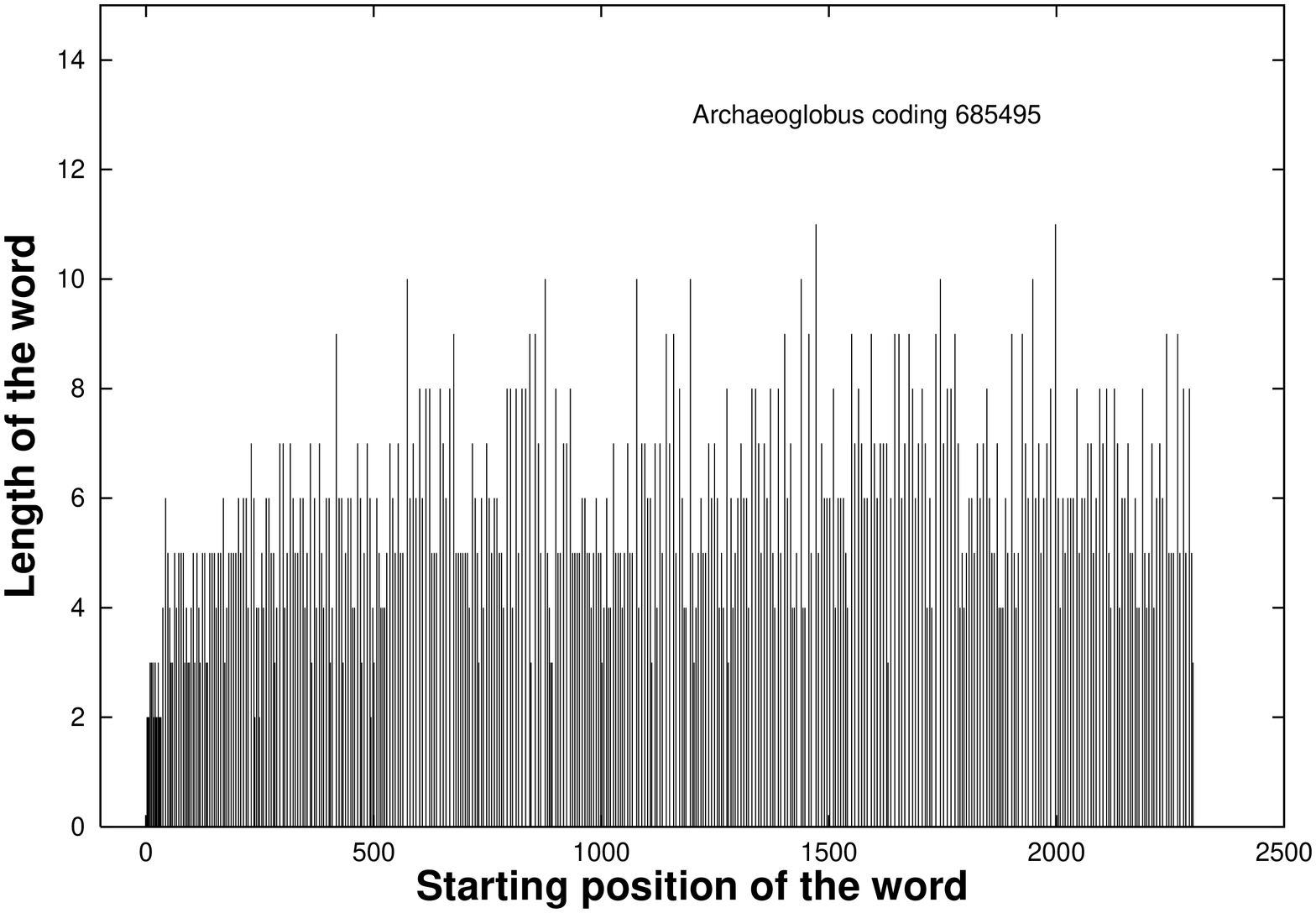,width=6.5cm,angle=270}}
&\raggedleft{(b)\psfig{figure=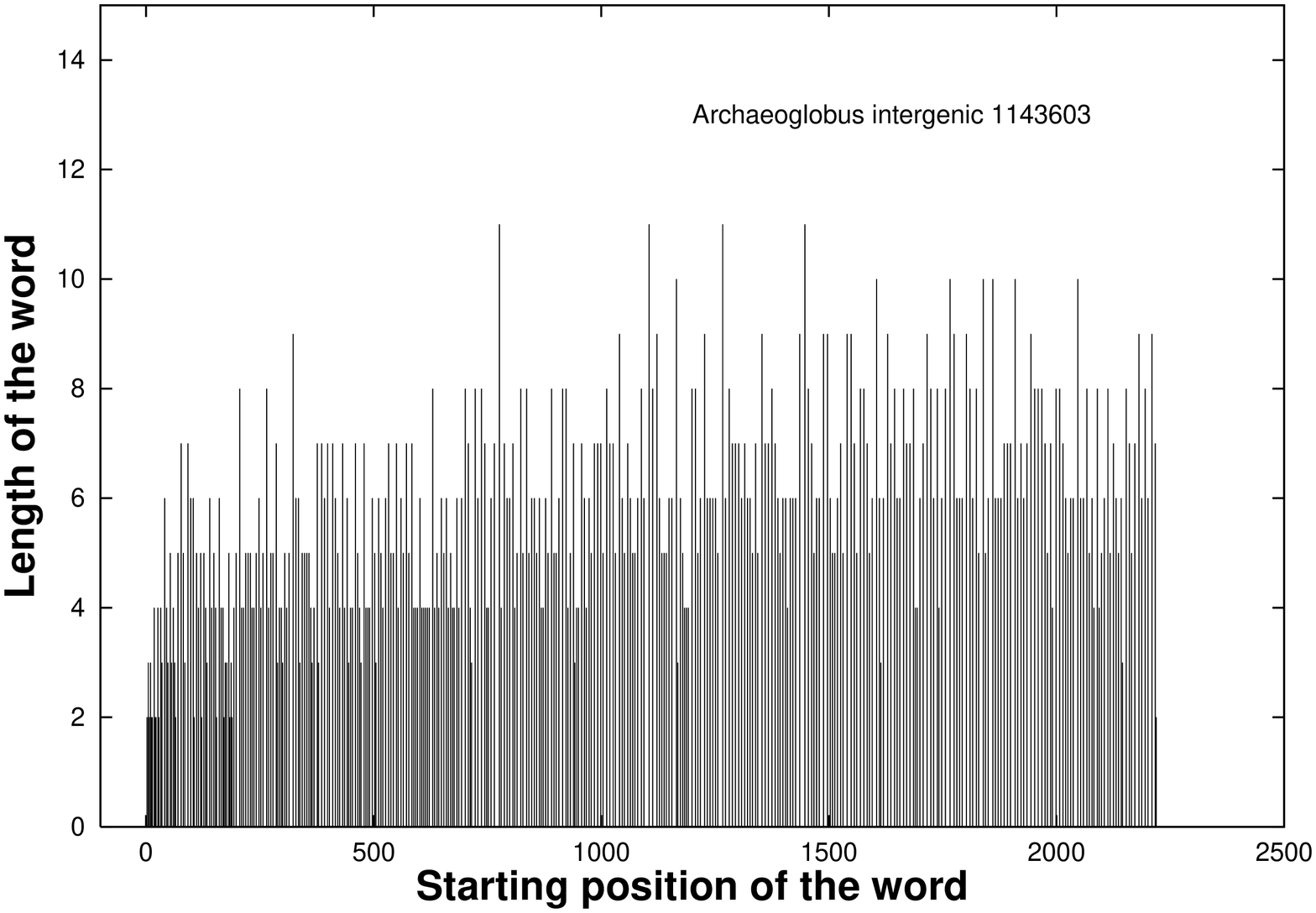,width=6.5cm,angle=270}}
\end{tabular}
\begin{tabular}{lr}
\raggedright{(c)\psfig{figure=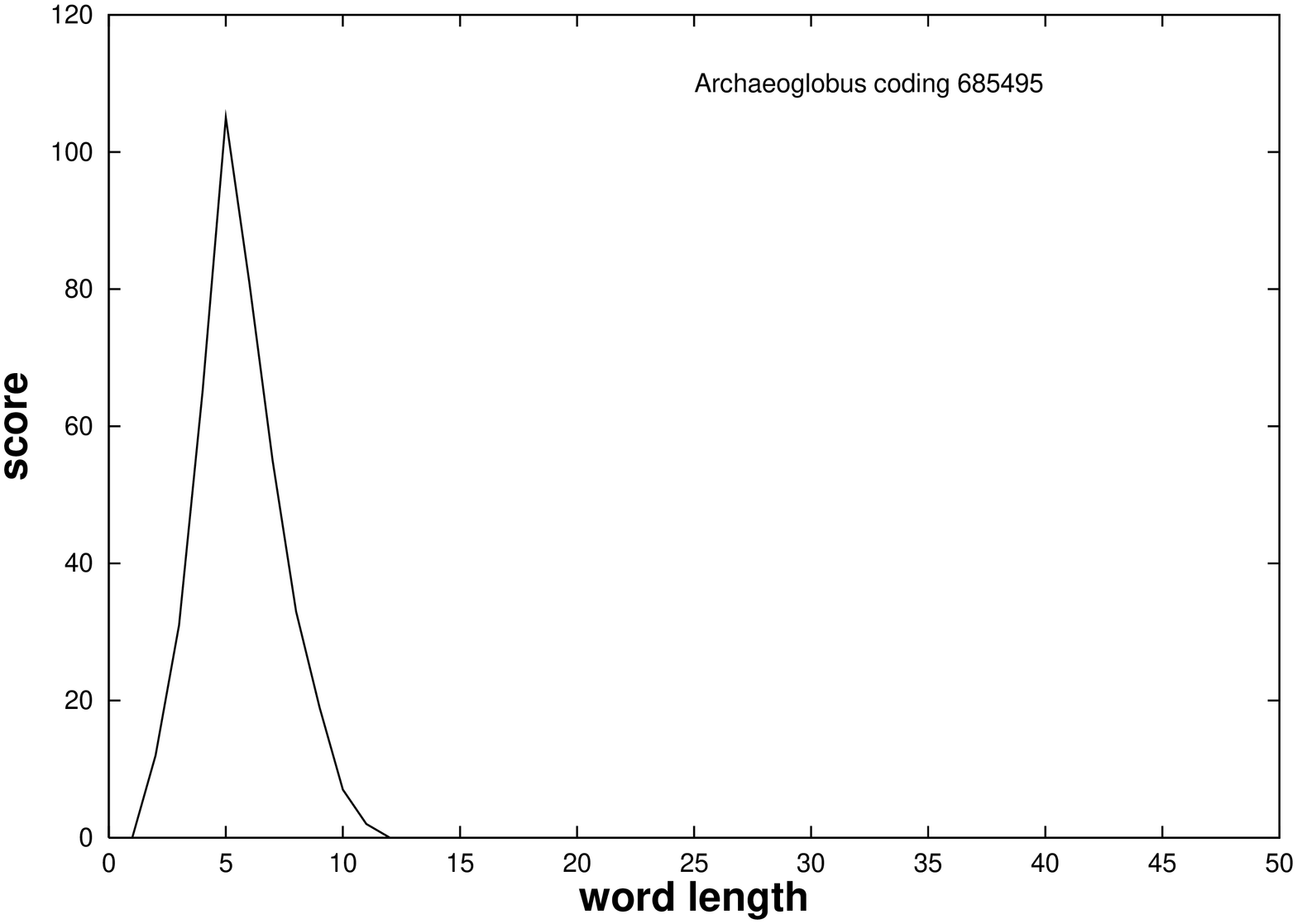,width=6.5cm,angle=270}}
&\raggedleft{(d)\psfig{figure=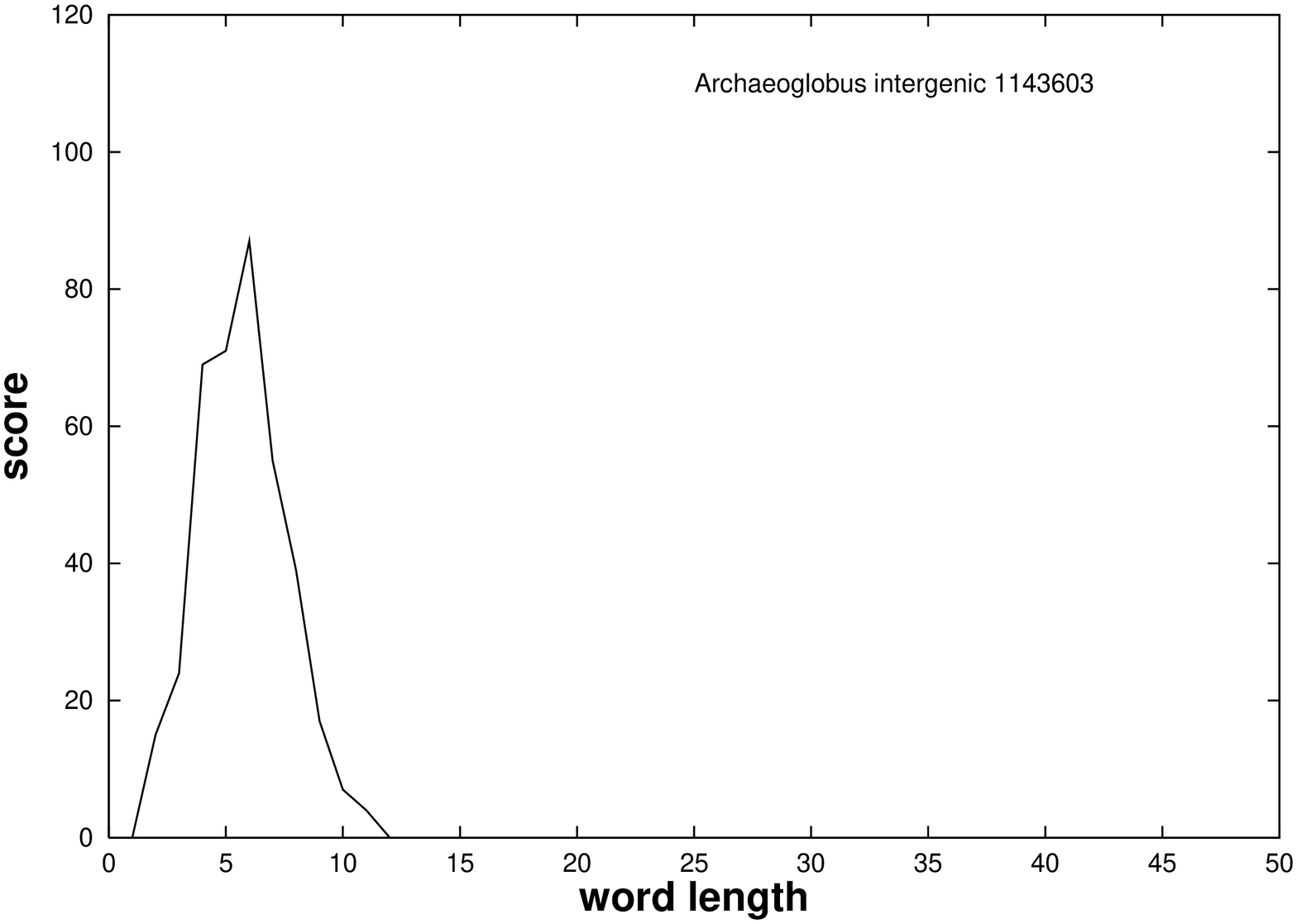,width=6.5cm,angle=270}}
\end{tabular}
\caption{\it Archaeoglobus fulgidus genome. Plots (a) and (b) show the
  location and length of the phrases in the parsing by the algorithm
  CASToRe, in non-atypical regions $Coding\_685495$ and
  $Inter\_1143603$, respectively. Graphs (c) and (d) illustrate the
  distribution of phrase length in the same
  regions.}\label{globNonatyp}
\end{figure}
First, we have plotted the length of the phrases in the dictionary
together with their position in the input sequence (see Figure
\ref{globNonatyp} $(a)$ and $(b)$). In both non-atypical regions, the
phrases are short and the maximal length is 11 bp. The Gaussian
distribution of phrase length confirms that these regions are not
regular, but highly variable (see Figure \ref{globNonatyp} $(c)$ and
$(d)$). The extent of the dictionary is great in both non-atypical
regions: 415 phrases in the dictionary of region $Coding\_685495$ and
393 phrases in the dictionary of region $Inter\_1143603$.

However, in the case of region $Coding\_685495$, the algorithm CASToRe
recognised 31 codons as phrases that are also used as prefix or suffix
words quite frequently. Table \ref{globcodon} illustrates the details
of this feature that has been found only in coding regions; in fact,
in non-atypical noncoding regions the codons that are recognised as
phrases are always a few.
\begin{table}\begin{center}
\begin{tabular}{|c|c|c||c|c|c|}
\hline 
\mbox{Codon}&\mbox{$\#$ prefix}&\mbox{$\#$ suffix}&\mbox{Codon}&\mbox{$\#$ prefix}&\mbox{$\#$ suffix}\\
\hline\hline
AAA&10&4&CTG&8&7\\
\hline
AAG&1&8&GCA&3&2\\
\hline
AAT&4&1&GCC&8&6\\
\hline
ACA&10&4&GCT&5&3\\
\hline
ACC&4&7&GGA&2&4\\
\hline
ACG&4&1&GGT&2&2\\
\hline
ACT&7&6&TAA&5&8\\
\hline
ATG&3&0&TAG&0&1\\
\hline
ATT&4 &8&TAT&2&4\\
\hline
CAA&14&10&TCA&6&6\\
\hline
CAG&3&7&TCC&8&2\\
\hline
CAT&0&0&TCT&7&5\\
\hline
CCG&3&1&TGT&2&6\\
\hline
CCT&2&3&TTA&5&7\\
\hline
CGG&0&0&TTG&0&4\\
\hline
CGT&5&7& & &\\
\hline
\end{tabular}
\caption{\it 31 different codons have been recognised as phrases in the
  parsing by the algorithm CASToRe, in region $Coding\_685495$ of the
  genome of {\it Archaeoglobus fulgidus}. Some of them have been also
  used as prefix or suffix of other phrases. Columns named $\#\ 
  prefix$ and $\#\ suffix$ indicate how many times the phrase has been
  used as a prefix or suffix.}\label{globcodon}
\end{center}
\end{table}
\begin{figure}
\begin{tabular}{lr}
\raggedright{(a)\psfig{figure=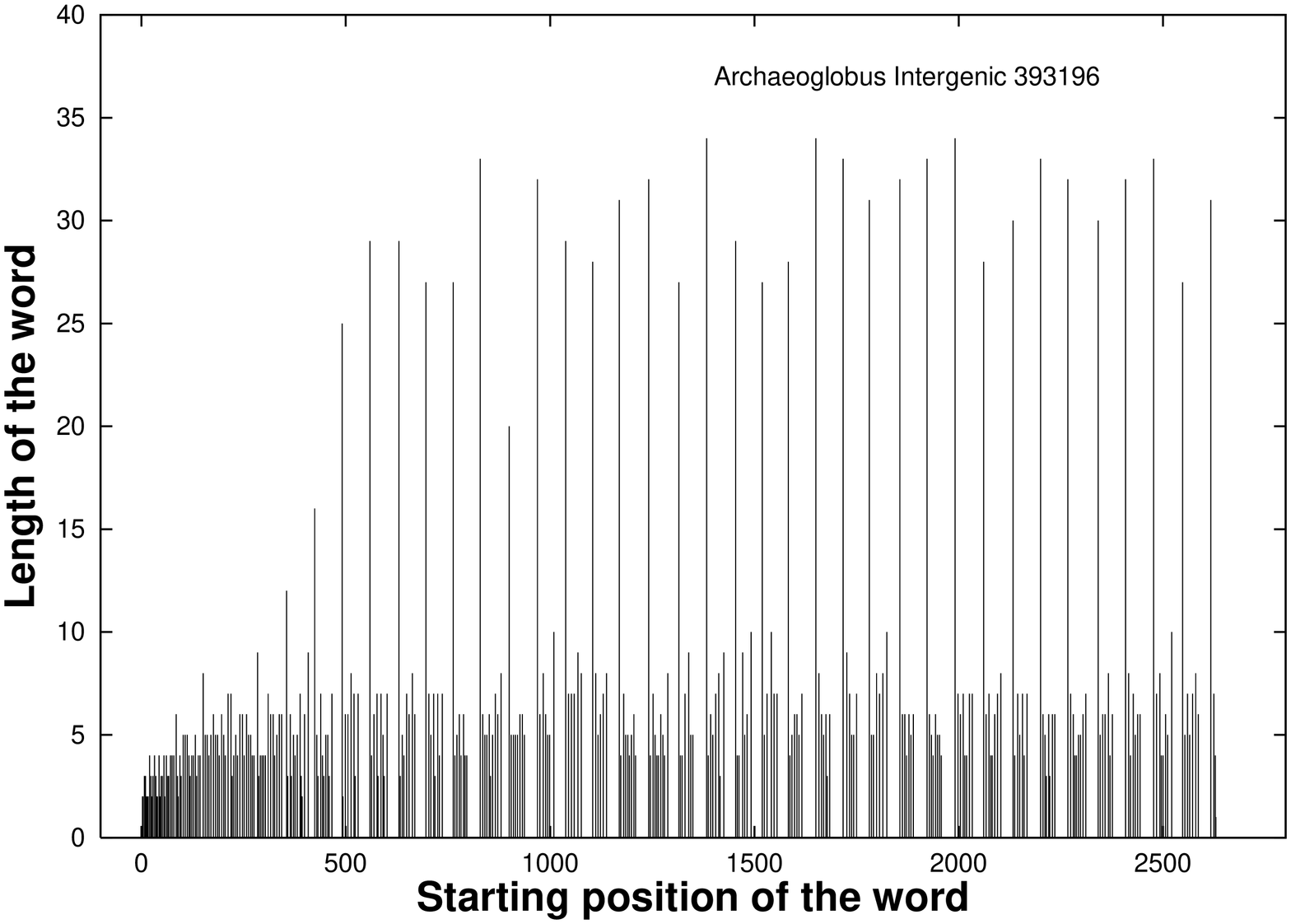,width=6.5cm,angle=270}}&
\raggedleft{(b)\psfig{figure=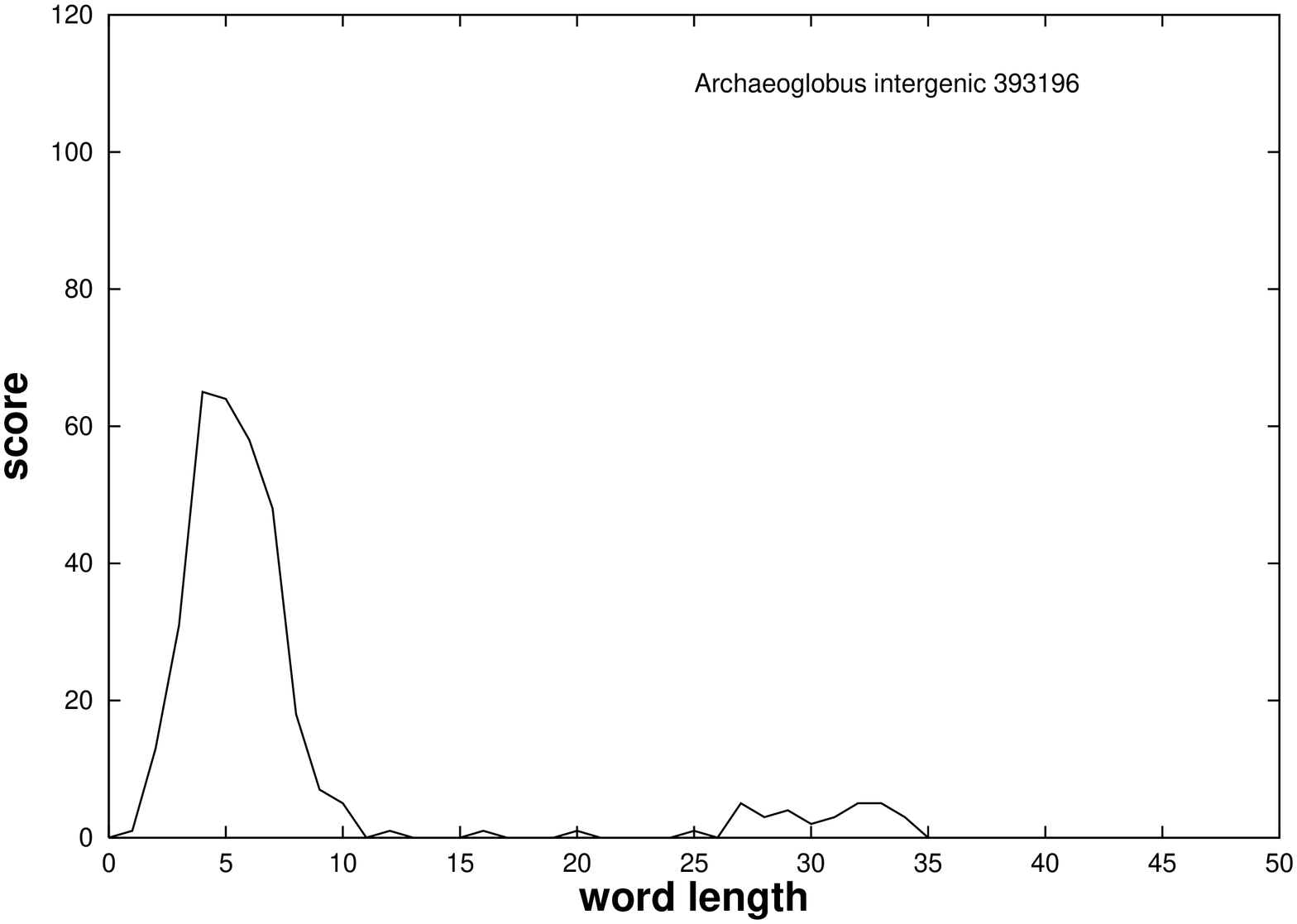,width=6.5cm,angle=270}}
\end{tabular}
\caption{\it Archaeoglobus fulgidus genome. Plot (a) shows
  the location and length of the phrases in the parsing by the
  algorithm CASToRe, in atypical region $Inter\_393196$. Graph (b) 
  illustrates the distribution of phrase length in the same
  region.}\label{globatyp}
\end{figure}

Conversely, the dictionary relative to fragment $Inter\_393196$, which
is atypical noncoding, shows completely different
characteristics. First of all, the dictionary contains 349
phrases. Moreover, Figure \ref{globatyp} $(a)$ shows that in this
sequence there should be recurrences of similar patterns, because of
the several long phrases (that is, longer than 25 bp) that are spread
along the whole sequence. Another feature, which will be paradigmatic
of atypical regions, is the anomalous (non-Gaussian) tail in the
distribution of phrase length (see Figure \ref{globatyp} $(b)$). The
distribution is no longer peaked at only one value, but there is a
significant occurrence of long words that could not be found in
non-atypical regions and is consistent with the presence of regularity
within any atypical region.

According to the dictionary obtained by means of algorithm CASToRe,
there is a dominant motif $\mathcal M$ of length 25 bp (phrase
nr. $109$ in the dictionary), that is also used 9 times as a prefix
and 3 times as a suffix.  Table \ref{tabMglobus} illustrates what the
dominant motif $\mathcal M$ and its descent are. We recall that the
descent of a phrase $\phi$ is the set of other phrases in the
dictionary such that $\phi$ is either their prefix or suffix or both.
\begin{table}\begin{center}
\begin{tabular}{c}
$\mathcal M=$AATCCCATTTTGGTCTGATTTCAAC\\
Descent of $\mathcal M\  :$\\
{\bf AATCCCATTTTGGTCTGATTTCAAC}ACA\\
{\bf AATCCCATTTTGGTCTGATTTCAAC}AG\\
{\bf AATCCCATTTTGGTCTGATTTCAAC}CAA\\
{\bf AATCCCATTTTGGTCTGATTTCAAC}CT\\
{\bf AATCCCATTTTGGTCTGATTTCAAC}GA\\
{\bf AATCCCATTTTGGTCTGATTTCAAC}GT\\
{\bf AATCCCATTTTGGTCTGATTTCAAC}TATTT\\
{\bf AATCCCATTTTGGTCTGATTTCAAC}TT\\
{\bf AATCCCATTTTGGTCTGATTTCAAC}TTTC\\
CCCTTTC{\bf AATCCCATTTTGGTCTGATTTCAAC}\\
CTTTC{\bf AATCCCATTTTGGTCTGATTTCAAC}\\
TTTC{\bf AATCCCATTTTGGTCTGATTTCAAC}\\
\end{tabular}
\caption{\it dominant motif $\mathcal M$ and its descent in atypical
  region $Inter\_393196$ of the genome of Archaeoglobus fulgidus.}
\label{tabMglobus}
\end{center}
\end{table}

The presence of a dominant motif partially motivates the many
oscillations in the $CIC$ growth, as depicted in Figure \ref{regLow}.
Furthermore, a complete explanation lays on the fact that the motif
$\mathcal M$ is also a sliding pattern in many other phrases (see
Table \ref{slidinglobus}). This is
an irrefutable evidence of the fact that this atypical region shows a
{\it variable periodicity} represented by the recurrence of the
motif $\mathcal M$ sometimes slightly modified, as in the case of
approximate repeats. 

Even if the biological usefulness of the motif $\mathcal M$ is still
unknown, another hint to its peculiarity is provided by the
compression of region $Inter\_393196$ by means of algorithm
$LZ77$. The motif $\mathcal M$ is a motif also in the dictionary
extracted by $LZ77$. Therefore, the idea that this motif should have a
precise biological meaning is even more convincing. Furthermore, this
example suggests that also approximate repeats generated by insertions may
be identified via $CIC$ method.
\begin{table}\begin{center}
\begin{tabular}{c}
$\mathcal M=$AATCCCATTTTGGTCTGATTTCAAC\\
Motif as a sliding pattern in:\\
TTC{\bf AATCCCATTTTGGTCTGATTTCAAC}\\
{\bf AATCCCATTTTGGTCTGATTTCAAC}GAAG\\
{\bf AATCCCATTTTGGTCTGATTTCAAC}CTCC\\
{\bf AATCCCATTTTGGTCTGATTTCAAC}TATTT\\
CTTTC{\bf AATCCCATTTTGGTCTGATTTCAAC}\\
CCTTTC{\bf AATCCCATTTTGGTCTGATTTCAAC}\\
TCTTTC{\bf AATCCCATTTTGGTCTGATTTCAAC}\\
TTC{\bf AATCCCATTTTGGTCTGATTTCAAC}TCC\\
TTTC{\bf AATCCCATTTTGGTCTGATTTCAAC}CTT\\
CGCTTTC{\bf AATCCCATTTTGGTCTGATTTCAAC}\\
{\bf AATCCCATTTTGGTCTGATTTCAAC}GAGGCGT\\
CCCTTTC{\bf AATCCCATTTTGGTCTGATTTCAAC}\\
CTCCTTTC{\bf AATCCCATTTTGGTCTGATTTCAAC}\\
CCTTTC{\bf AATCCCATTTTGGTCTGATTTCAAC}TA\\
ACTTTC{\bf AATCCCATTTTGGTCTGATTTCAAC}AG\\
TTTC{\bf AATCCCATTTTGGTCTGATTTCAAC}TTTA\\
CTTTC{\bf AATCCCATTTTGGTCTGATTTCAAC}ATC\\
GTCTCTTTC{\bf AATCCCATTTTGGTCTGATTTCAAC}\\
CACGCTTTC{\bf AATCCCATTTTGGTCTGATTTCAAC}\\
ACCCCTTTC{\bf AATCCCATTTTGGTCTGATTTCAAC}
\end{tabular}
\caption{\it phrases where the motif $\mathcal M$ is a sliding
  pattern. The motif is written bold typed.}
\label{slidinglobus}
\end{center}
\end{table}
\subsection{Methanococcus jannaschii}
\begin{figure}
\centerline{\psfig{figure=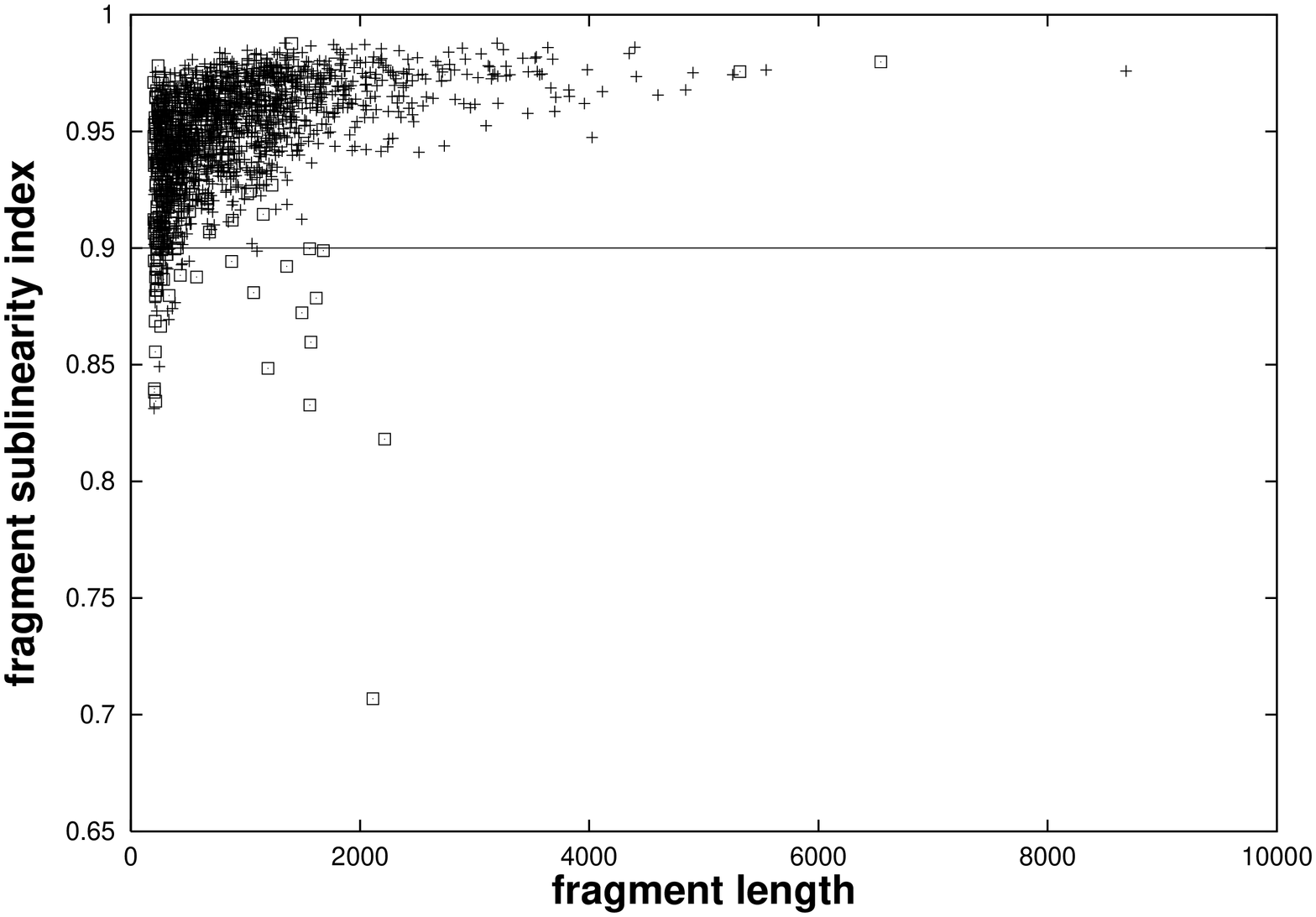,width=8cm,angle=270}}
\caption{\it Methanococcus jannaschii genome. Of each functional
  region, its length and the corresponding sublinearity index are
  plotted. The crosses ($+$) are referred to coding regions, while the
  squares ($\square$) are referred to intergenic regions.  The
  horizontal line is the threshold for the sublinearity index, under
  which the region is atypical. }\label{LGmjann}
\end{figure}

{\it Methanococcus jannaschii} is a thermophilic (48-94$^\circ$ C),
strict anaerobic Archaebacterium living at pressures of over 200
atmospheres. It is an autotroph which gets its energy from hydrogen
and carbon dioxide producing methane and it is capable of nitrogen
fixation. Morphologically, it is characterized by having two bundles
of flagella at the same cellular pole. The genome of {\it
Methanococcus jannaschii} consists of the main circular chromosome and
two circular extrachromosomal elements (ECE), one large and one
small. We have analysed only the main chromosome.

In this genome we shall show one atypical region, whose sublinearity
index is particularly low and having approximately the same extent as
the other regions that have been already analysed. However, as it
is shown on Figure \ref{LGmjann}, this genome presents many other long
atypical regions, that will be studied in future work.

The atypical region we have analysed is
\begin{itemize}
\item $Inter\_236189$: atypical region, length $L=2112\ bp$,
  sublinearity index \linebreak$\mathcal{G}_{_Z}=0.707$, fragment
  complexity $K= 1.405$.
\end{itemize}
\begin{figure}
\centerline{\psfig{figure=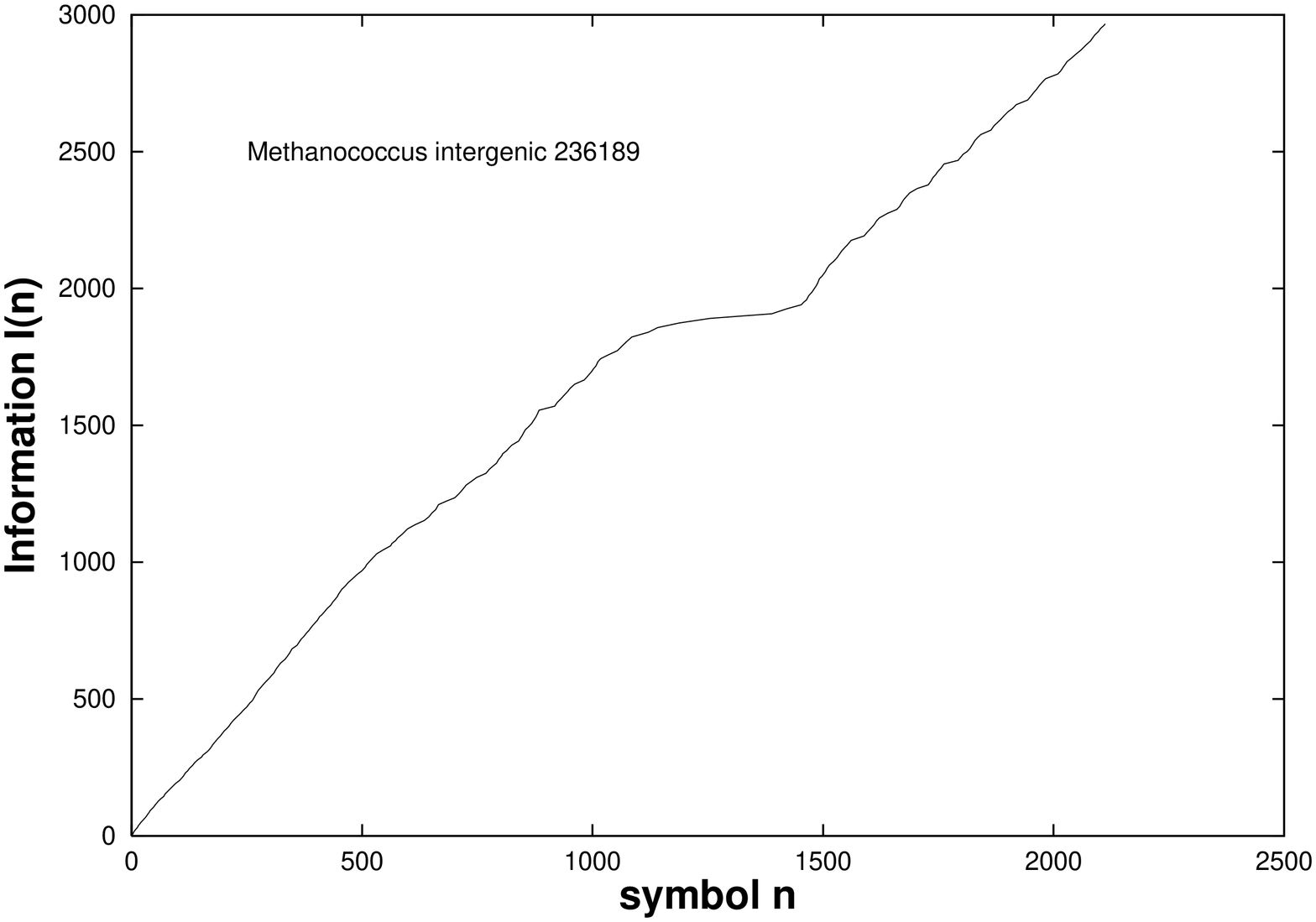,width=8cm,angle=270}}
\caption{\it Methanococcus jannaschii genome. The behaviour of the
  information content of region $Inter\_236189$ grows sublinearly with
  index 0.707. The picture is in linear scale.}\label{mjannLow}
\end{figure}
The behaviour of the information content in atypical region
$Inter\_236189$ is twofold: until the first 1500 base pairs have been
encoded, the growth is almost logarithmic, while in the final part the
$CIC$ increase is faster (see Figure \ref{mjannLow}). Therefore, the
first part of the sequence should be more regular than the second one.

\begin{figure}
\begin{tabular}{lr}
\raggedright{(a)\psfig{figure=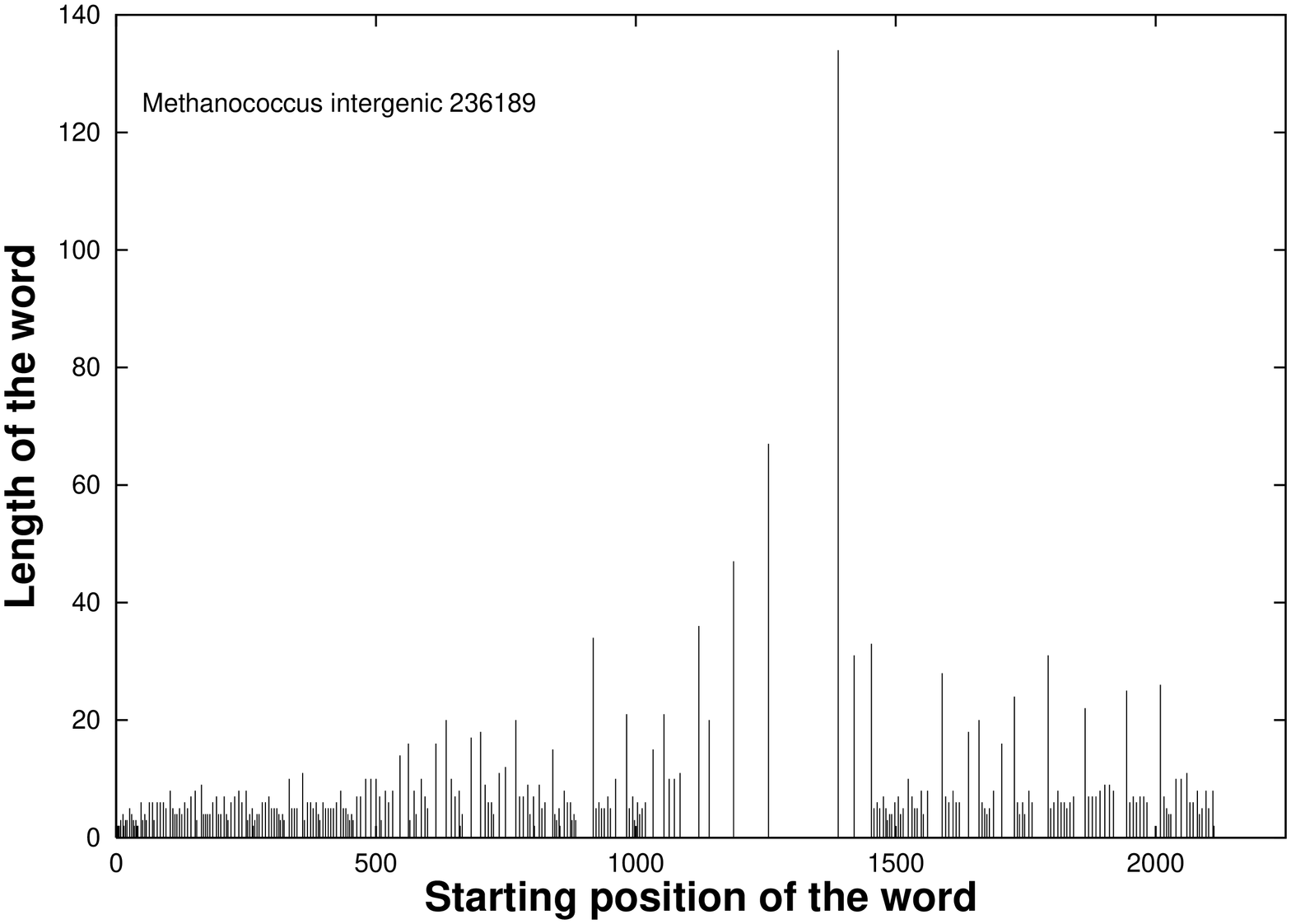,width=6.5cm,angle=270}}&
\raggedleft{(b)\psfig{figure=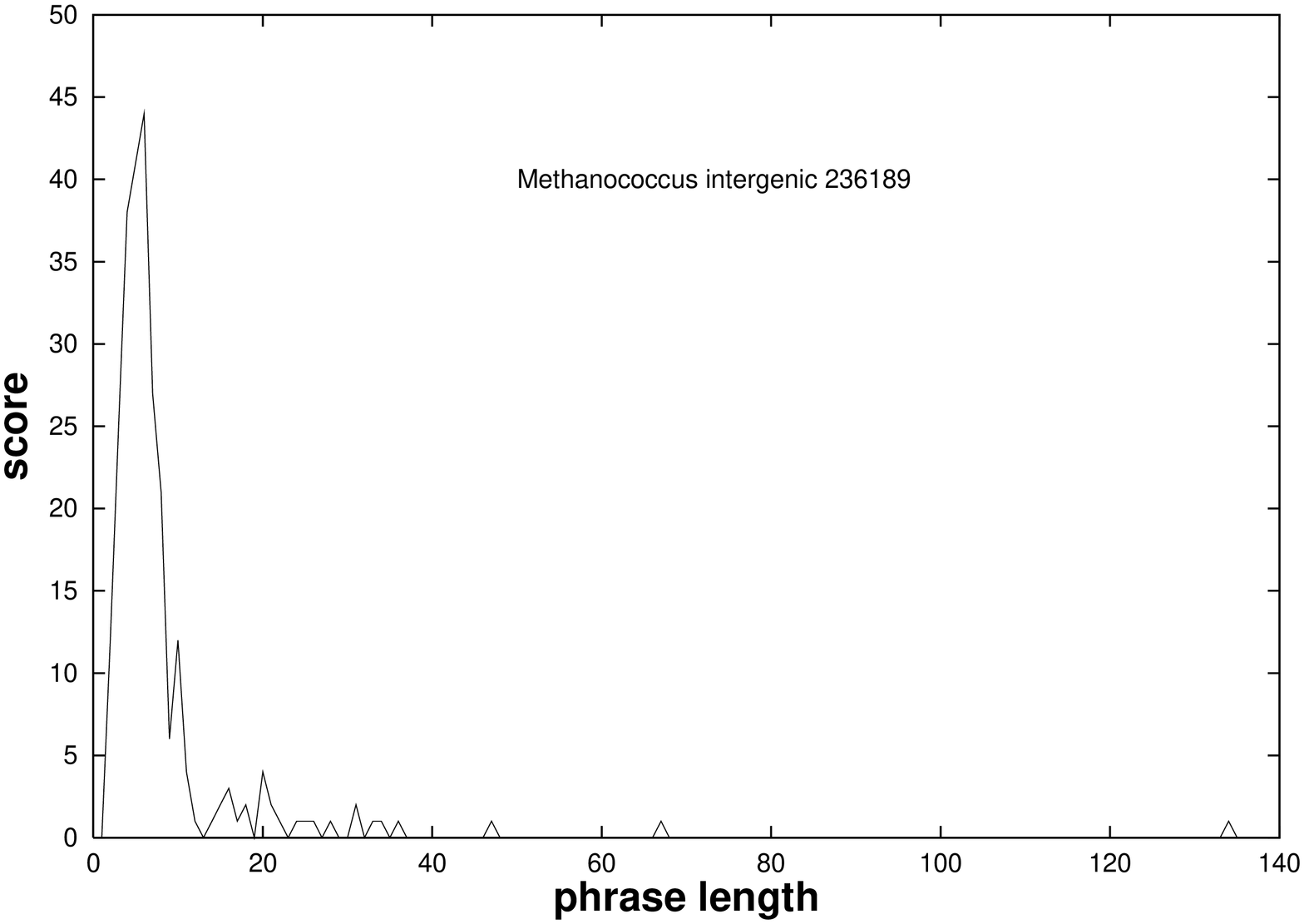,width=6.5cm,angle=270}}
\end{tabular}
\caption{\it Methanococcus jannaschii genome. Plot $(a)$ shows 
  location and length of the phrases in the parsing by the algorithm
  CASToRe of region $Inter\_236189$. In graph (b) the corresponding
  distribution of phrase length is pictured.}\label{mjannparole}
\end{figure}

This aspect is well-represented in graph $(a)$ of Figure
\ref{mjannparole}. The presence of longer and longer phrases before
1500 bp have been compressed is an evidence for the existence of
highly repetitive subsequences in the first half, whereas in the
second half of the input sequence $Inter\_236189$ the previous
regularity is broken and only brief repetitions can be
found. Consequently, the extent of the dictionary is low: there are
only 264 phrases. 

As in the case of the analysed atypical region of genome of {\it
  Archaeoglobus fulgidus}, the distribution of phrase length has an
anomalous non-Gaussian tail that comprehends also a phrase that is 134
  bp long (Figure \ref{mjannparole} $(b)$).

\begin{table}\begin{center}
\begin{tabular}{|l|}
\hline
AATTAAAATCAGACCGTTTCGGAATGGAAAT\\
\hline
AGACCGTTTCGGAATGGAAAT\\
\hline
AGACCGTTTCGGAATGGAAATGAT\\
\hline
AGGGAACCCTAAAAAGGTTC\\
\hline
AGGGAACCCTAAAAAGGTTCCCTTGAGGGTT\\
\hline
AGGGAACCCTAAAAAGGTTCCCTTGAGGGTTCATTAAAATCAGACCGTT\\
                                       TCGGAATGGAAATCTGTT\\
\hline
AGGGAACCCTAAAAAGGTTCCCTTGAGGGTTCATTAAAATCAGACCGTT\\
TCGGAATGGAAATCTGTTAGGGAACCCTAAAAAGGTTCCCTTGAGGGTT\\
CATTAAAATCAGACCGTTTCGGAATGGAAATCTGTT\\
\hline
ATTAAAATCAGACCGTTTCGGAATGGAAATGATT\\
\hline
CATTAAAATCAGACCGTTTCGGAATGGAAATTC\\
\hline
CATTAAAATCAGACCGTTTCGGAATGGAAATCTGTT\\
\hline
CCTTGAGGGTTCATTAAAATCAGACCGTTTCGGAATGGAAATCTGTT\\
\hline
GTATTAAAATCAGACCGTTTCGGAAT\\
\hline
GTTTCGGAATGGAAATCTGTT\\
\hline
GTTTCGGAATGGAAATGAAT\\
\hline
GTTTCGGAATGGAAATGATT\\
\hline
GTTTCGGAATGGAAATTTTT\\
\hline
TAAAATCAGACCGTTTCGGAAT\\
\hline
TAAAATCAGACCGTTTCGGAATGGAAAT\\
\hline
TAAAATCAGACCGTTTCGGAATGGG\\
\hline
\end{tabular}
\caption{\it Methanococcus jannaschii genome. Phrases longer than 20
  bp are listed, coming from the dictionary relative to atypical
  region $Inter\_236189$.}\label{tabmjann}
\end{center}
\end{table}

For what concerns the analysis of recurrent phrases in the dictionary,
it holds that only phrases that are shorter than 10 bp are used more
than three times as prefix word or suffix word. As it is shown in
Table \ref{tabmjann}, the phrases longer than 20 bp (that correspond
to the high ``spikes'' of Figure \ref{mjannparole} $(a)$) do not allow a
dominant motif to be determined in such a definite way as in the case
of atypical region $Inter\_393196$ of {\it Archaeoglobus fulgidus}
genome. The increasingly longer phrases that have been detected in
graph \ref{mjannparole} $(a)$ are not generated by coupling
the prefix word to itself (as it would have been if there were a precise
periodicity), but prefix and suffix words were different from each
other and neither they are subsequent. Again, the longest phrases
coincides with the longest ones found by means of the algorithm $LZ77$.

However, the main point of distinction of this atypical region is that
all long phrases are rich in T$^n$A$^m$-patterns. This fact, together
with the positive homology response classify this region as a promoter
region containing a subregion known as {\it TATA box}. The promoter
sequence could be located using program PROSCAN Version 1.7
(\cite{promoter}).

The dictionary of this region provides another example of regularity
in DNA sequences, different from the one coming from the genome of
{\it Archaeoglobus fulgidus}.

\subsection{Arabidopsis thaliana}
{\it Arabidopsis thaliana} is a small flowering plant that is widely
used as a model organism in plant biology. {\it Arabidopsis} is a member of
the mustard (Brassicaceae) family, which includes cultivated species
such as cabbage and radish. {\it Arabidopsis thaliana} is the first
plant for which the complete genome has been sequenced. Its genome
consists of five chromosomes, but we have analysed only chromosomes II
and IV. Since the research regarding this genome is
still {\it in itinere}, here we shall present some very preliminar results
concerning chromosome II.

\begin{figure}
\begin{tabular}{lr}
\raggedright{(a)\psfig{figure=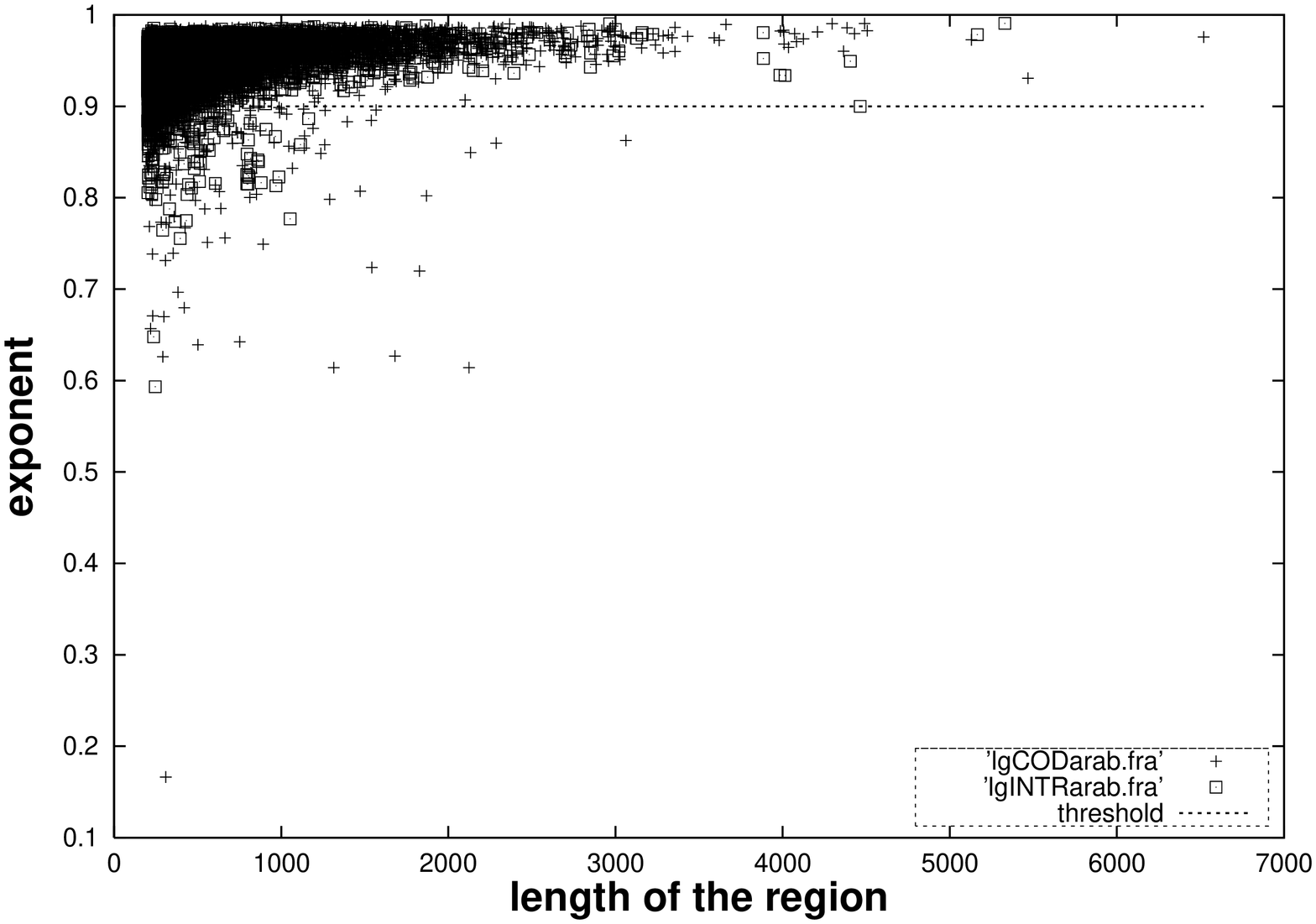,width=6.5cm,angle=270}}&
\raggedleft{(b)\psfig{figure=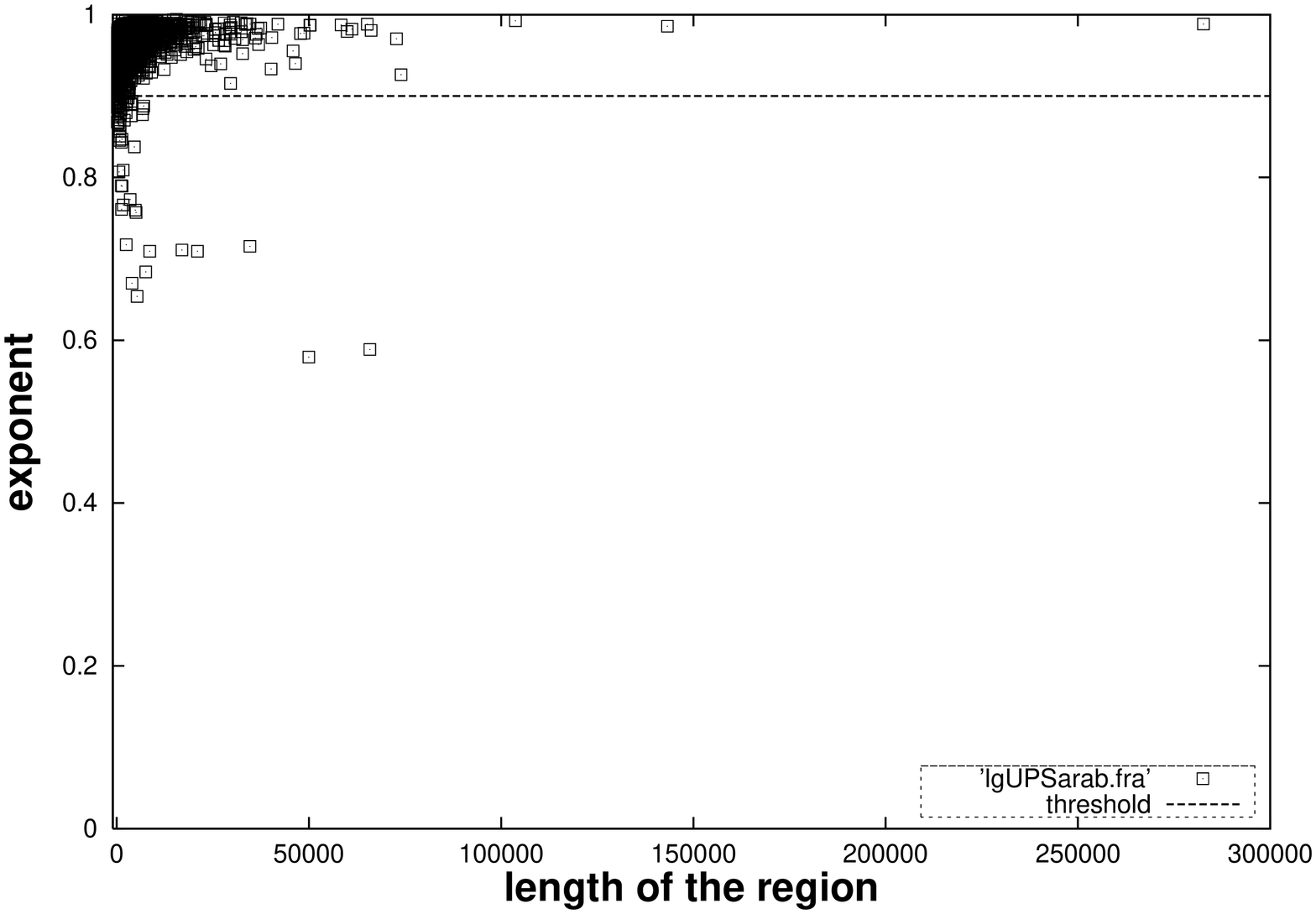,width=6.5cm,angle=270}}
\end{tabular}
\caption{\it Arabidopsis thaliana genome. Of each functional region, its
  length and the corresponding sublinearity index are plotted. In
  picture (a), the crosses ($+$) are referred to coding regions, while
  the squares ($\square$) are referred to introns. In picture (b), the
  squares ($\square$) are referred to intergenic regions. In both
  plots, the horizontal line is the threshold for the sublinearity
  index, under which the region is atypical. }\label{lgarab}
\end{figure}

The atypical regions we have analysed are
\begin{itemize}
\item $Coding\_8330271$: atypical region, length $L=309\ bp$,
  sublinearity index $\mathcal{G}_{_Z}=0.166$, fragment
  complexity $K= 1.113$.
\item $Inter\_22564763$: atypical region, length $L=65849\ bp$,
  sublinearity index $\mathcal{G}_{_Z}=0.589$, fragment
  complexity $K= 0.911$.
\end{itemize}
These regions have been chosen as peculiar among the
many atypical regions (see Figure \ref{lgarab}) belogning to this
genome: a short and very regular coding region and a long intergenic
region.
\begin{figure}
\centerline{(a)\psfig{figure=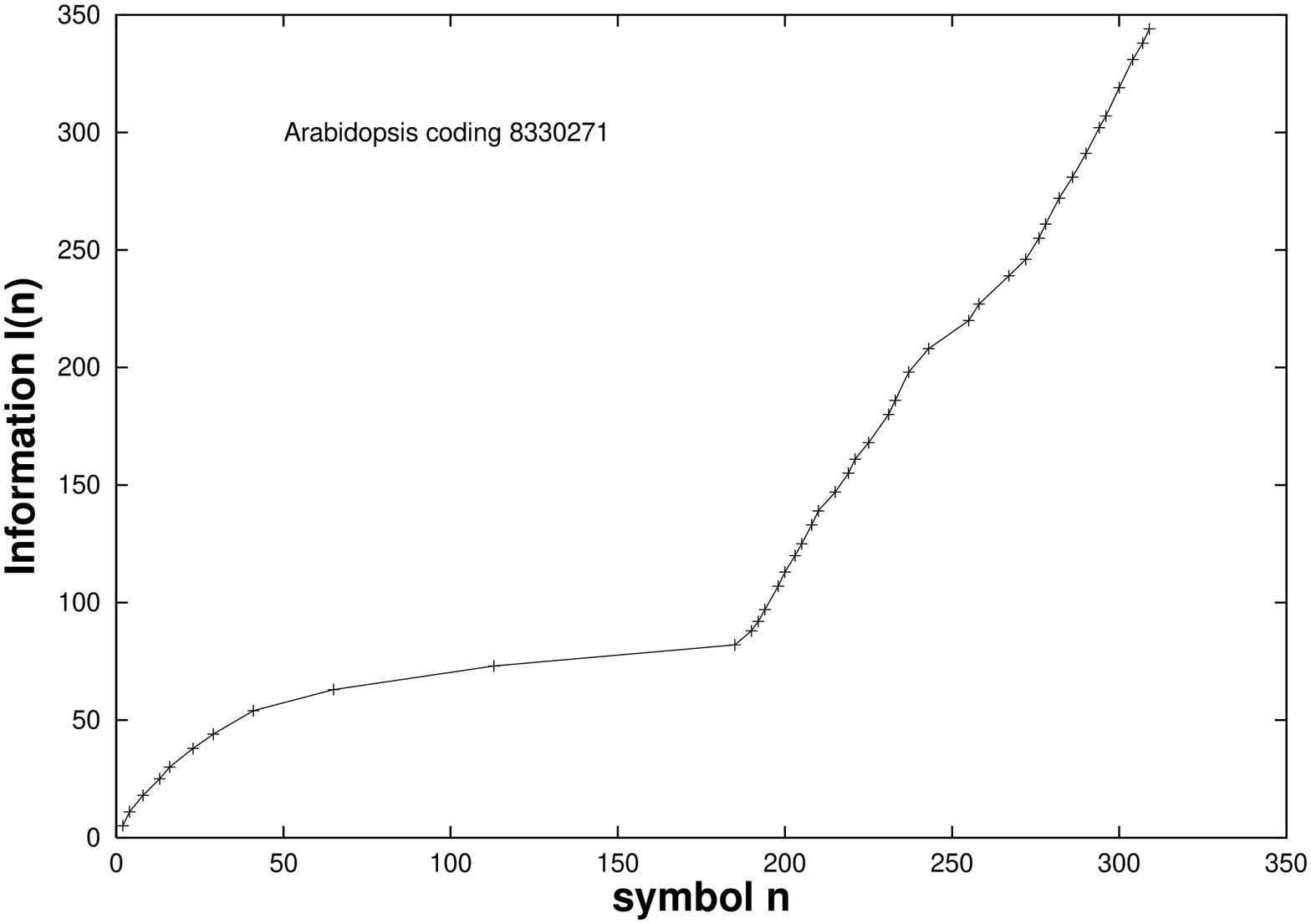,width=7cm,angle=270}}
\begin{tabular}{lr}
\raggedright{(b)\psfig{figure=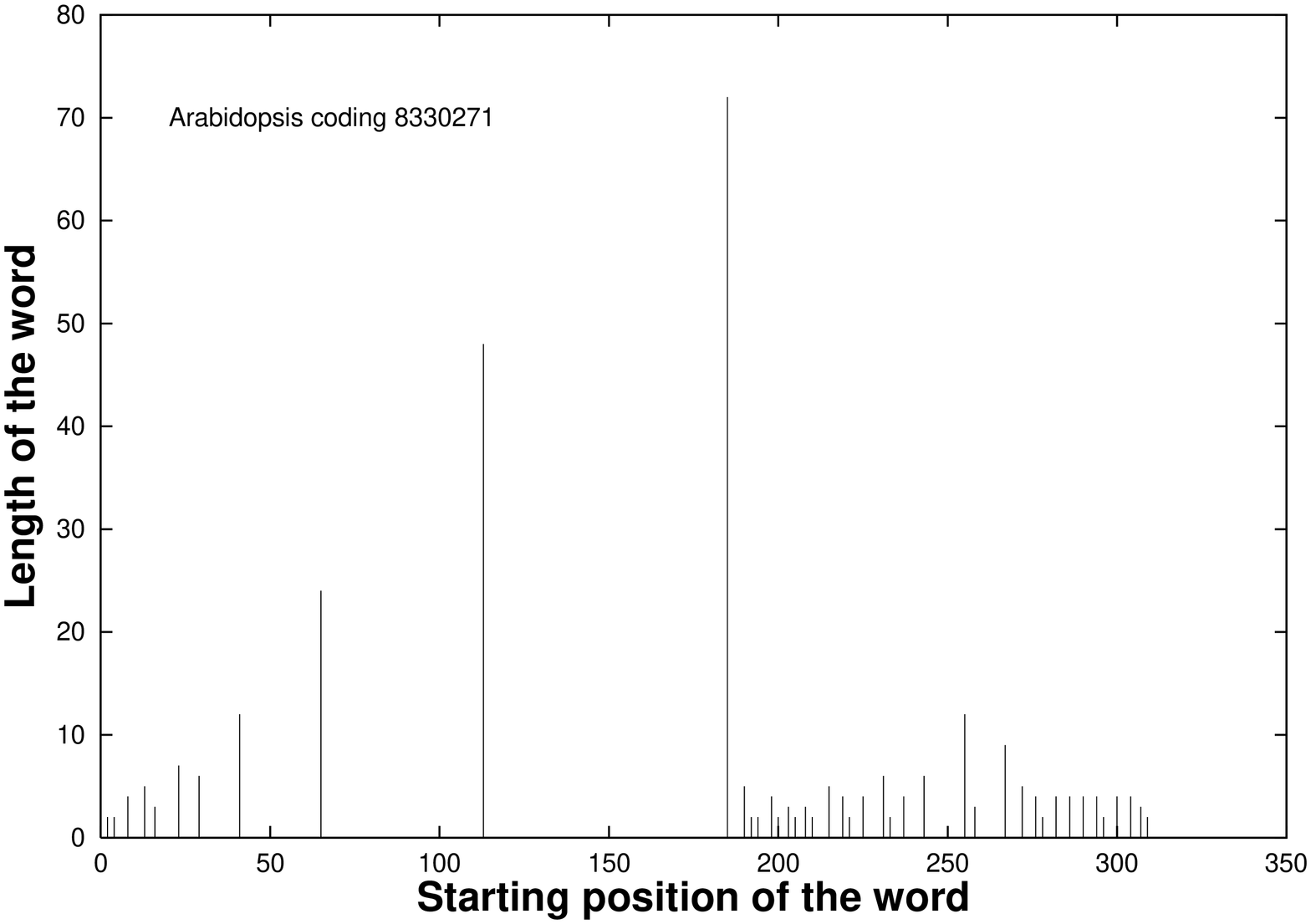,width=6.5cm,angle=270}}&
\raggedleft{(c)\psfig{figure=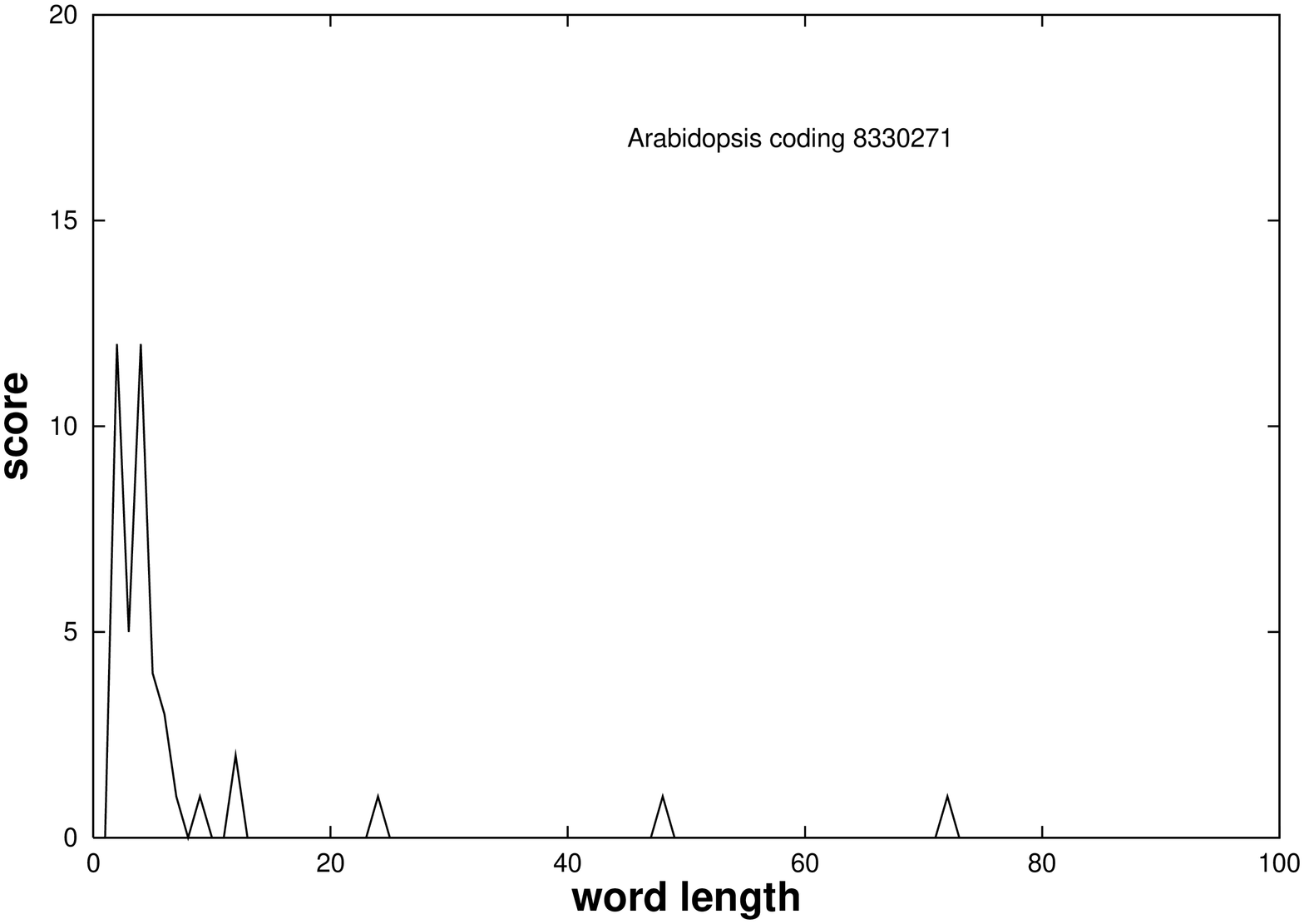,width=6.5cm,angle=270}}
\end{tabular}
\caption{\it Atypical region $Coding\_8330271$. (a) The Information
Content growth is logarithmic for the main part of the sequence. The
word length doubling is shown on plot (b) and the multimodal
distribution of word length is illustrated in (c).}\label{proteinarab}
\end{figure}

The atypical region $Coding\_8330271$ is characterized by a period
$^\prime GA^{\ \prime}$ that is repeated for most part of the sequence
(the first 200 bp). This is made evident both from the $I(n)$ plot on
Figure \ref{proteinarab} $(a)$, which is definitely logarithmic in the
first part, and from the word length doubling highlighted in Figure
\ref{proteinarab} $(b)$. Also, the multimodal distribution of word
length reflects the atypical nature of this regions, while the maximal
length is $12$ bp, which confirms that the characteristic maximal
length in non-atypical coding regions is about $11-12$ bp (for
instance, see \ref{globNonatyp} (c)). The putative protein that may be
obtained by translating this coding region is following protein
Atg219370:
$$
\begin{array}{l}
\mathrm{ERERGSERERERERERERERERERERERERERERERER}\\
\mathrm{EREREREREREREREREREREREREKHKPATLAKNRRR}\\
\mathrm{RFVKNRRRRDHRRRISIIDGYESQF*V}\\
\end{array}
$$ 

In the above notation, each letter corresponds to an amino acid, while
the star indicates the end of the protein. This putative protein is
very rich in Glutamate (E) and Arginine (R), but its function is still
unknown and consideration should be given to the fact that the actual
existence of this protein in the living organism has not yet been
confirmed by biomolecular laboratory experiments, therefore this
fragment has been classified as coding onyl by means of statistical
predicitive methods .
\begin{figure}
\begin{tabular}{lr}
\raggedright{(a)\psfig{figure=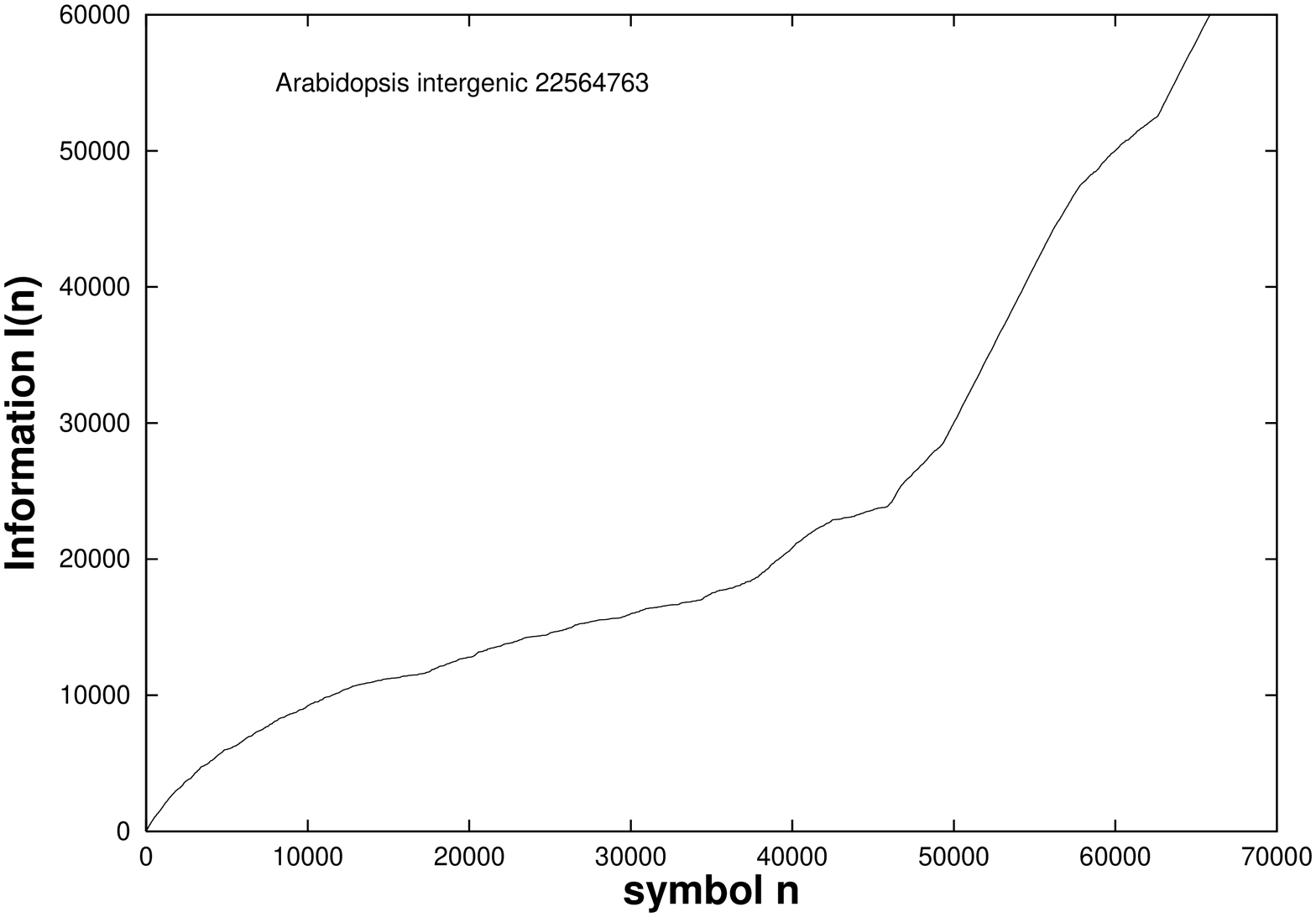,width=6.5cm,angle=270}}&
\raggedleft{(b)\psfig{figure=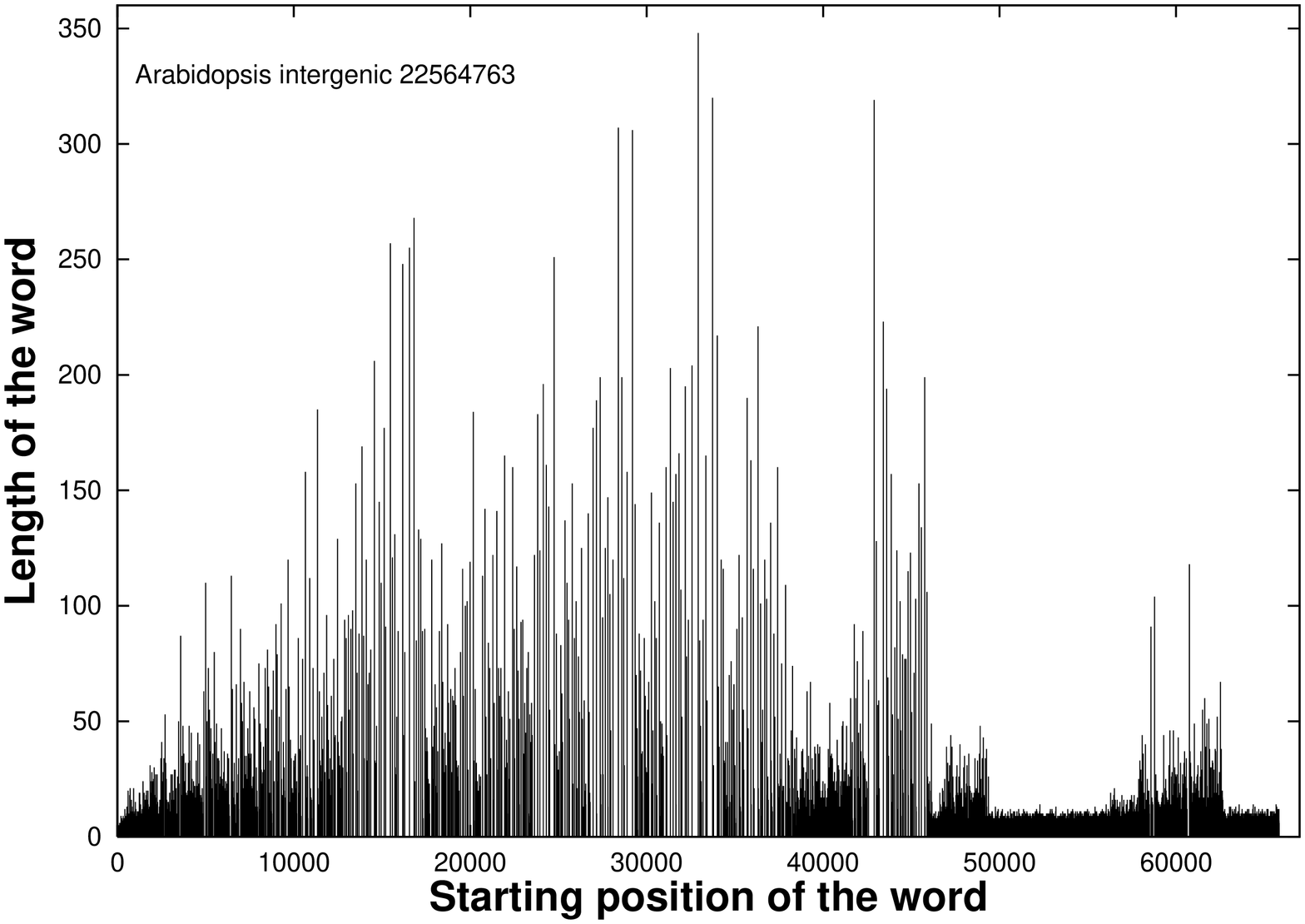,width=6.5cm,angle=270}}
\end{tabular}
\begin{tabular}{lr}
\raggedright{(c)\psfig{figure=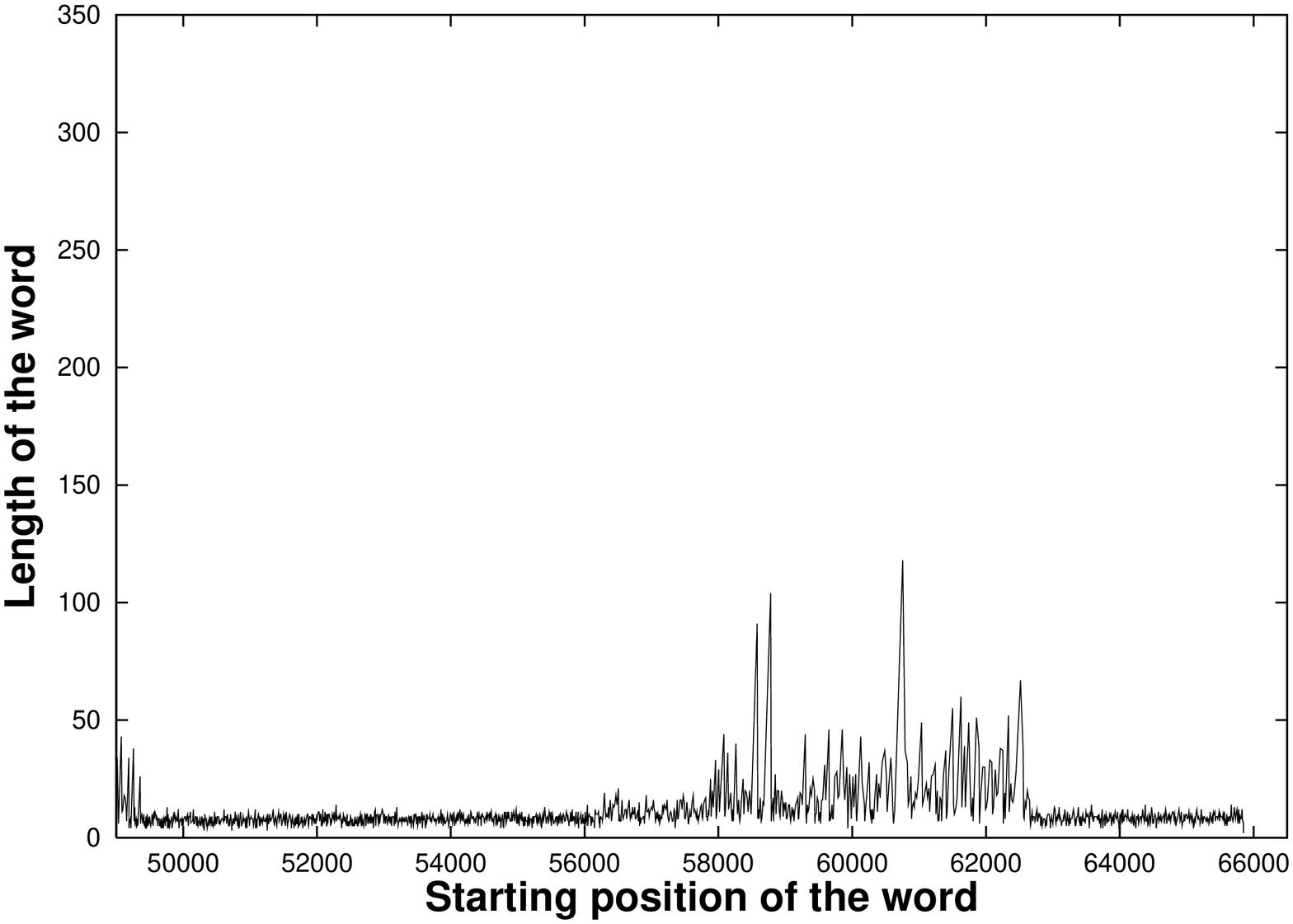,width=6.5cm,angle=270}}&
\raggedleft{(d)\psfig{figure=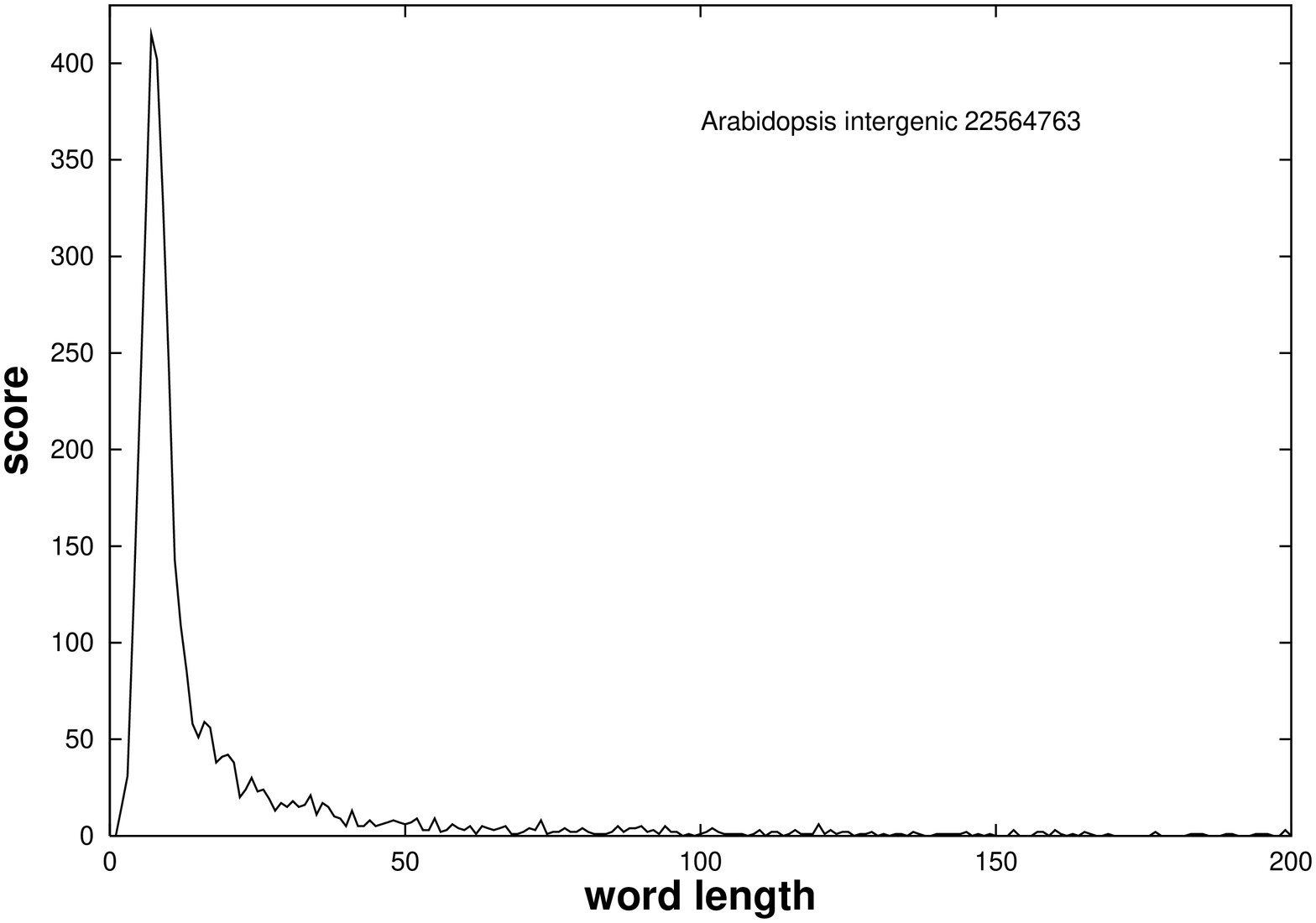,width=6.5cm,angle=270}}
\end{tabular}
\caption{\it Arabidopsis thaliana genome (chromosome II). (a) The
behaviour of Information Content of atypical region $Inter\_22564763$
grows in a very peculiar way. Its sublinearity index has been
evaluated as 0.589. (b) The plot shows location and length of the
phrases in the parsing obtained by the algorithm CASToRe. (c) The plot
is an enhancement of the final part of the atypical region
$Inter\_22564763$. (d) The distribution of phrase length for the
aforementioned parsing is pictured.}\label{arabgene}
\end{figure}

The atypical region $Inter\_22564763$ was a challenging task, because
not only the Information Content growth shows an abrupt change around
$50000$ bp (Figure \ref{arabgene} $(a)$), but also the word length is
subjected to a deep decrease when reaching that threshold, although at
that point the dictionary already contained more than $1700$ phrases,
most of them longer than 50 bp (Figure \ref{arabgene} $(b)$ and
$(c)$).

It was this twofold look of the region that suggested that in the
final part of this region (from $50000$ bp to $65849$ bp) there might
have been some coding sequences. This was also supported by the
prevailing length of about $11-12$ bp, which, as it was already
pointed out, may be considered as characteristic of coding regions.
\begin{figure}
\centerline{\psfig{figure=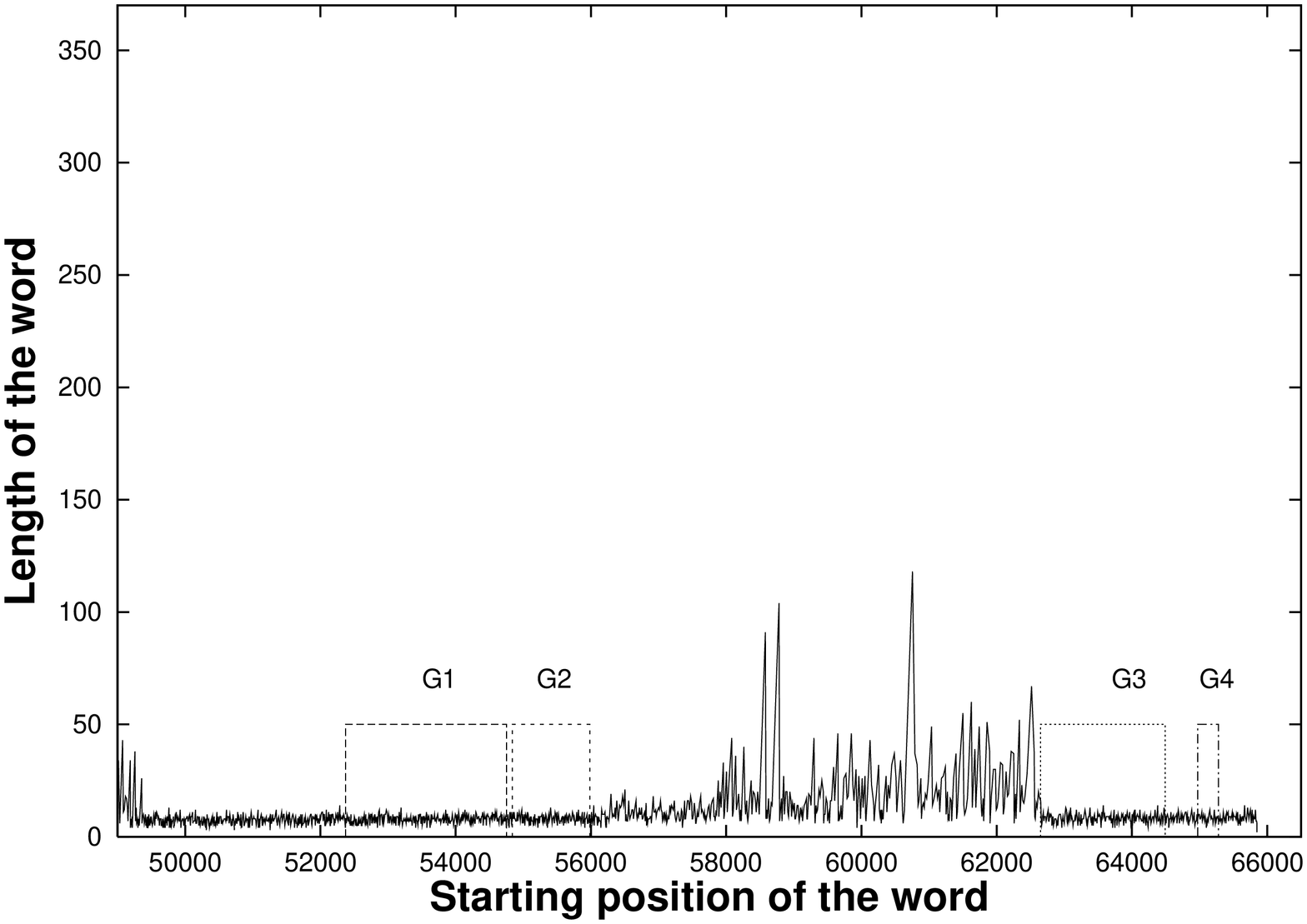,width=12cm,angle=270}}
\caption{\it Arabidopsis thaliana genome (chromosome II). Same part of
atypical region $Inter\_22564763$ as plot (c) in Figure
\ref{arabgene}. The boxes correspond to the location of the
four predicted genes (labelled as $^\prime G1^\prime,^\prime
G2^\prime,^\prime G3^\prime,^\prime G4^\prime$) as they have been
predicted looking for similarities with Arabidopsis thaliana known
genes.}\label{cfrarabgene}
\end{figure}
As a result, four putative genes G1, G2, G3 and G4 have
been located by means of Hidden Markov Model-based program
FGENESH\footnote{This program is available at the website
www.softberry.com to which we refer concerning the reliability and
efficiency of the algorithm.} that has been created for predicting
multiple genes and their structure in genomic DNA sequences. The
analysis via FGENESH has been exploited with respect to known genes in {\it
Arabidopsis thaliana}. Their predicted position is illustrated in Figure
\ref{cfrarabgene}.

%
\section{Final remarks and future work}
We have shown that complete genomes may be analysed in some of their
distinctive features by means of the Computable Information Content
obtained via compression algorithms. The Information Content may be
used to extract regions having an atypical information growth, which
is strictly connected to the presence of highly repetitive subregions
that might be supposed to have a regulatory function within the
genome. Different types of sublinearities have been associated to
different biogical features. These results shall pave the way for a
more profound understanding of the local compressibility of genomes
and for a more detailed identification of motifs and patterns that are
significant to some biological function, in view of a joint use
together with other predictive methods.

\end{document}